%% file: dyneos.tex
\documentclass[paper]{JHEP3}
\usepackage{epsfig}
\usepackage{graphicx}
\usepackage{amsmath}
\usepackage{multirow}
\usepackage{lscape}
\usepackage{ulem}

\vspace{2cm}
\def\lsi{\raise0.3ex\hbox{$<$\kern-0.75em\raise-1.1ex\hbox{$\sim$}}}
\def\gsi{\raise0.3ex\hbox{$>$\kern-0.75em\raise-1.1ex\hbox{$\sim$}}}
\newcommand{\lsim}{\mathop{\lsi}}
\newcommand{\gsim}{\mathop{\gsi}}

\newcommand{\be}{\begin{equation}}
\newcommand{\ee}{\end{equation}}
\newcommand{\dd}{\textmd{d}}

\long\def\symbolfootnote[#1]#2{\begingroup\def\thefootnote{\fnsymbol{footnote}}\footnote[#1]{#2}\endgroup}

\title{
The QCD equation of state with dynamical quarks}
\author{
Szabolcs~Bors\'{a}nyi$^a$, Gergely~Endr\H{o}di$^b$, Zolt\'{a}n~Fodor$^{a,b}$, Antal Jakov\'ac$^a$,
S\'{a}ndor~D.~Katz$^b$, Stefan Krieg$^{a,c}$, Claudia~Ratti$^a$ and K\'{a}lm\'{a}n~K.~Szab\'o$^a$\\
$^a$Department of Physics, University of Wuppertal, Gauss 20, D-42119, 
Germany\\
$^b$Institute for Theoretical Physics, E\"otv\"os University, P\'azm\'any
1, H-1117 Budapest, Hungary\\
$^c$Center for Theoretical Physics, MIT, Cambridge, MA 02139-4307, USA
}

\preprint{WUB/10-12, MIT-CTP 4168}

\abstract{ 
The present paper concludes our investigation on the QCD equation of state with
$2+1$ staggered flavors and one-link stout improvement. We extend our previous
study [JHEP 0601:089 (2006)] by choosing even finer lattices.  Lattices with
$N_t=6,8$ and $10$ are used, and the continuum limit is approached by checking
the results at $N_t=12$. A Symanzik improved gauge and a stout-link improved
staggered fermion action is utilized. We use physical quark masses, that
is, for the lightest staggered pions and kaons we fix the $m_\pi/f_K$ and
$m_K/f_K$ ratios to their experimental values. The pressure, the interaction
measure, the energy and entropy density and the speed of sound are presented as
functions of the temperature in the range $100 \ldots 1000 \textmd{MeV}$. We
give estimates for the pion mass dependence and for the contribution of the
charm quark.  We compare our data to the equation of state obtained by the
``hotQCD'' collaboration. 
}


\begin{document}

\section{Introduction} \input{intro}

\section{\label{methods}Methods} \input{methods}
\subsection{\label{methods_act}Lattice discretization} \input{methods_act}
\subsection{\label{methods_lcp}Lines of constant physics} \input{methods_lcp}
\subsection{\label{methods_int}The integral technique} \input{methods_int}
\subsection{\label{methods_hrg}Hadron Resonance Gas Model} \input{methods_hrg}

\section{\label{result}Results} \input{result}

\section{Conclusions, outlook} \input{summary}

\acknowledgments{Computations were performed on the Blue Gene supercomputers at
FZ J\"ulich and on clusters at Wuppertal and also at the E\"otv\"os University,
Budapest. This work is supported in part by European Union (EU) grant I3HP;
Deutsche Forschungsgemeinschaft grants FO 502/2 and SFB-TR 55 and
(FP7/2007-2013)/ERC no. 208740, and by the U.S. Department of Energy under
Grant No. DE-FG02-05ER25681.}

\bibliographystyle{JHEP}
\bibliography{dyneos}

\appendix
\section{\label{app_spline}The pressure from a 2D spline fit} \input{app_spline}
\newpage
\section{\label{app_table}Tables} \input{app_table}

\end{document}

%% file: intro.tex
The study of QCD thermodynamics and that of the phase diagram are receiving
increasing attention in recent years.  A transition occurs in strongly
interacting matter from a hadronic, confined system at small temperatures and
densities to a phase dominated by colored degrees of freedom at large
temperatures or densities.  Lattice simulations indicate that the transition at
vanishing chemical potential is merely an analytic crossover
\cite{Aoki:2006we}. Even if strictly speaking there is no phase transition, it
is common to use the words confined and deconfined phases for the low and high
temperature regimes.  This field of physics is particularly appealing because
the deconfined phase of QCD can be produced in the laboratory, in the
ultrarelativistic heavy ion collision experiments at CERN SPS, RHIC at
Brookhaven National Laboratory, ALICE at the LHC and the future FAIR at the
GSI. The close interplay between experimental data, numerical simulations on
the lattice and phenomenological models offer the unique possibility of
understanding the properties of matter under extreme conditions.  The
experimental results available so far show that the hot QCD matter produced
experimentally exhibits robust collective flow phenomena, which are well and
consistently described by near-ideal relativistic hydrodynamics
\cite{Teaney:2000cw,Teaney:2001av,Kolb:2003dz}. These hydrodynamical models
need as an input an Equation of State (EoS) which relates the local
thermodynamic quantities. Therefore, an accurate determination of the QCD EoS
is an essential ingredient to understand the nature of the matter created in
heavy ion collisions, as well as to model the behavior of hot matter
in the early universe.

Numerical simulations of QCD thermodynamics on the lattice are reaching
unprecedented levels of accuracy, and a variety of data are now available for
the EoS, including works in the quenched approximation
\cite{Boyd:1996bx,Okamoto:1999hi}, two-flavor simulations
\cite{Bernard:1996cs,AliKhan:2001ek}, studies with heavier-than-physical
\cite{Karsch:2000ps,Kanaya:2009nq}, and almost physical quark masses
\cite{Bernard:2006nj,Cheng:2007jq,Bazavov:2009zn,Cheng:2009zi}.  In Reference
\cite{Aoki:2005vt}, our collaboration has presented results for the EoS of 2+1
flavor QCD with physical quark masses, on lattices with temporal extensions
$N_t=4$ and $N_t=6$ and for temperatures up to $3T_c$. Data for an EoS
involving physical masses together with a careful continuum limit are so far
missing; their relevance for the physics of the Quark-Gluon Plasma (QGP) is
obvious.

An issue that is receiving increasing attention in recent years, is whether the
charm quark can give an important contribution to the QCD EoS, in the range of
temperatures which are reached in heavy ion collisions. It is often assumed
that it can be neglected, the charm mass being too heavy to play any role at
$T\simeq 2-3T_c$. However, perturbative QCD predicts that its contribution to
thermodynamic observables is relevant at surprisingly low temperatures, down to
$T\simeq 350$ MeV \cite{Laine:2006cp}. Recent exploratory lattice studies have
confirmed these expectations \cite{Cheng:2007wu,Levkova:2009gq}, indicating a
non-negligible contribution of the charm quark to thermodynamics already at
1.2-2$T_c$.  These results have been obtained on rather coarse lattices
($N_t=4,~6$) and with the charm quark treated in the quenched approximation.

Most of the available results on the QCD EoS have been obtained using improved
staggered fermion actions. This formulation does not preserve the flavor
symmetry of continuum QCD; as a consequence, the spectrum of low lying hadron
states is distorted.  Recent analyses performed by various collaborations
\cite{Huovinen:2009yb,Huovinen:2010tv,Borsanyi:2010bp} have pointed out that
this distortion can have a dramatic impact on the thermodynamic quantities. In
order to quantify this effect, one can compare the low temperature behavior of
the observables obtained on the lattice, to the predictions of the Hadron
Resonance Gas (HRG) model, and monitor how it reaches the continuum limit (for
nonvanishing lattice spacings the spectrum is distorted, which has an influence
on the prediction of the HRG model).

In this paper, we present our most recent results for several thermodynamic
observables: pressure, energy density, entropy density, trace anomaly and speed
of sound, for a system of $n_f=2+1$ flavors of dynamical quarks. We also
determine the contribution of the charm quark to the pressure of the system.
The charm quark is treated at the partially quenched level, i.e. we use the
same gauge field configurations as in the $n_f=2+1$ case.  We improve our
previous findings \cite{Aoki:2005vt} by choosing finer lattices ($N_t=8,~10$
and a few checkpoints at $N_t=12$). We work again with physical light and
strange quark masses: we fix them by reproducing the physical ratios
$m_\pi/f_K$ and $m_K/f_K$ for the lightest staggered tastes of these mesons and
by this procedure \cite{Aoki:2009sc} we get $m_s /m_{u,d} = 28.15$. Several
values for the charm quark mass are used in the range $m_c/m_s=10.75\ldots 20$.
As we will see, the different sets of data corresponding to different $N_t$
nicely agree with each other for all observables under study: for this reason,
we expect that discretization effects are tiny. We also check that there are no
significant finite size effects in the lattices that we use, by performing a
set of simulations in a box with a size of 7 fm at the transition temperature.
Our results are obtained in the range 100 MeV $\lsim T\lsim$ 1000 MeV. The
simulations are performed by using the tree-level Symanzik improved gauge, and
stout-improved staggered fermion action that we already used in
\cite{Aoki:2005vt}. Recently it has become clear
\cite{Huovinen:2009yb,Huovinen:2010tv,Borsanyi:2010bp} that the taste splitting
due to the staggered formulation affects dramatically several thermodynamical
observables at low temperatures.  Together with the projected smeared links
used in the recently proposed HISQ action \cite{Bazavov:2009mi,Bazavov:2010sb},
the stout-smearing \cite{Morningstar:2003gk} has the smallest taste violation
among the ones used so far in the literature for large scale thermodynamical
simulations. Other staggered fermion actions with larger taste violation, such
as the ``asqtad'' and ``p4fat'' actions used by the ``hotQCD'' collaboration,
suffer from these lattice artefacts more seriously. It is therefore not
surprising that the results obtained in the present paper are rather different
from the ones obtained by the ``hotQCD'' collaboration in Reference
\cite{Cheng:2009zi}.

The paper is divided into two major parts: Section \ref{methods} describes the
methods and techniques, which were used in this study. Those, who are
interested in the results only, can skip this part and can go directly to
Section \ref{result}, where we present and analyze our lattice results.

%% file: methods.tex
The fundamental quantity of finite temperature field theory is the
partition function, any thermodynamic observable can be derived from it. After
integrating out the quark fields ($\psi_q$, with flavor index $q$) it can be
written as a path integral over the gauge field $U$:
\be
\mathcal{Z}= \int [dU]
\exp\left(-\beta S_g(U) \right) \prod_q \left( \det M(U, m_q) \right)^{1/4}.
\ee
Here $S_g$ is the gauge action, $\beta$ is related to the gauge coupling as
$\beta=6/g^2$, the staggered quark Dirac operator is $M$ and $m_q$ is the mass of the
quark with flavor $q$. In this paper we use the same mass for the up and down
flavors ($m_{ud}$), the mass of the strange quark is denoted by $m_s$ and the mass of
the charm quark by $m_c$. The free energy density is $f = -\frac{T}{V}
\log \mathcal{Z}$, where $T$ denotes the temperature and $V$ the three-volume
of the system. In the thermodynamic limit the pressure is related to the free
energy density as
\be
p = -\lim_{V\to\infty}f.
\label{eq:press}
\ee
In the following we will always assume to have a large enough, homogeneous
system, so that the pressure can always be identified with the negative of the
free energy. Later on we will check, that this condition can be safely assumed
in case of the lattice simulations. Having calculated the pressure as a 
function of the temperature $p(T)$, all other thermodynamic observables can
also be reconstructed. The trace anomaly $I=\epsilon-3p$
is a straightforward derivative of the normalized pressure:
\be
I=T^5 \frac{\partial}{\partial T} \frac{p(T)}{T^4}
\label{eq:Idef}
\ee
Using the pressure and the trace anomaly the energy density $\epsilon$, the
entropy density $s$ and the speed of sound $c_s$ can be calculated as
\be
\epsilon= I + 3p,\quad\quad s = \frac{\epsilon+p}{T},\quad\quad c_s^2=\frac{\dd p}{\dd \epsilon}.
\label{eq:eosq}
\ee

Next we will describe in detail the way we have calculated the thermodynamic
observables in Equations (\ref{eq:press}), (\ref{eq:Idef}), (\ref{eq:eosq}) on the
lattice. Our choice for the discretization of the gauge action and the quark
Dirac operators and a discussion of the discretization effects is presented in
Subsection \ref{methods_act}. On the lattice, the gauge coupling and the quark
masses are not independent, their relation is dubbed Lines of Constant Physics
(LCP) and is described in Subsection \ref{methods_lcp}. The pressure can be
determined from lattice observables by an integral, this technique, with our
specific improvements, is presented in Subsection \ref{methods_int}. Finally we
describe the HRG model in Subsection \ref{methods_hrg}, which
is used for comparison with the lattice calculations in the low temperature
regime. 

%% file: methods_act.tex
The compact Euclidean spacetime of temperature $T$ and three-volume $V$ is
discretized on a hypercubic lattice with $N_t$ and $N_s$ points in the temporal
and spatial directions: 
\be
\label{eq:t}
T = \frac{1}{N_t a},\quad\quad V = (N_s a)^3,
\ee
where $a$ is the lattice spacing. At a fixed $N_t$, the temperature can be set
by varying the lattice spacing. This implies varying the bare parameters
of the lattice action accordingly (see Subsection \ref{methods_lcp}).
At a fixed temperature, lattice discretization effects can be investigated by
changing $N_t$, we used $N_t=6,8,10$ and $12$ in this study. Finite volume
effects were also studied by considering two different spatial volumes in
the case of $N_t=6$. For the renormalization we took lattices with $N_t\ge
N_s$, at the present precision they can all be considered as having zero
temperature. The following table summarizes the lattice sizes ($N_t\times N_s^3$) and the temperature ranges of this study:
\begin{equation}
\begin{array}{|c|c|c|c|}
\hline
N_t & {\rm finite ~} T & {\rm zero ~ } T & T {\rm ~ values}\\
\hline
6&6\times18^3, 6\times36^3 & 18\times18^3, 36\times18^3& 100\dots 1000 {\rm ~MeV}\\
\hline
8&8\times24^3  & 24\times24^3& 100\dots 1000 {\rm ~MeV}\\
\hline
10&10\times32^3 & 32\times32^3, 96\times32^3 &100\dots 365 {\rm ~MeV}\\
\hline
12&12\times32^3, 12\times64^3 & 32\times32^3, 64\times64^3 & 132, 167, 223 {\rm ~MeV}\\
\hline
\end{array}
\nonumber
\end{equation}

In this work we use the same action as in our earlier studies on the QCD
transition's order and characteristic temperatures
\cite{Aoki:2006we,Aoki:2005vt,Borsanyi:2010bp,Aoki:2009sc,Aoki:2006br}. As for
the gauge field, it means a tree level Symanzik improvement. The fermions are
discretized using the one-link staggered action with stout-smeared gauge links
\cite{Morningstar:2003gk}. We employ two levels of the analytic smearing, each
with the smearing parameter $\rho=0.15$. In order to get rid of the unwanted
doublers of the staggered formulation we use the ``rooting procedure''
\cite{Durr:2005ax}. 

The description of our updating algorithm can be also found in the above
papers. The effectivity of these algorithms towards the continuum limit has
been questioned recently \cite{Schaefer:2009xx}, where a serious increase in
the autocorrelation time of the topological charge was observed. On Figure
\ref{fig:charge} we show the Monte-Carlo time history of the topolological
charge in a zero temperature run on one of our finest lattice spacings ($a=
0.054$ fm). The charge is measured by the naive operator after performing 30
HYP smearing steps on the gauge configurations. In our setup no dangerous
critical slowing down can be observed. We have also looked for correlations
between the topological charge and various thermodynamical observables. Those
relevant for the equation of state show no significant dependence on the
topological sectors.

In lattice field theory one attempts to reduce the cutoff effects by the so
called improvement program. Not only the lattice action but also lattice
observables can be improved.  In lattice thermodynamics the cutoff effects
depend on $N_t$, for $N_t\to\infty$ they disappear.  In our analysis we use a
tree-level improvement for the pressure: we divide the lattice results with the
appropriate improvement coefficients. These factors can be calculated
analytically for our action and in case of the pressure we have the following
values on different $N_t$'s:
\begin{equation}
\begin{array}{|c|c|c|c|}
\hline
N_t=6&N_t=8&N_t=10&N_t=12\\
\hline
1.517&1.283&1.159&1.099\\
\hline
\end{array}
\nonumber
\end{equation}
We use the same factor for the two different spatial volumes in our finite
volume study ($6\times 18^3$ and $6\times36^3$).  Using thermodynamical
relations one can obtain these improvement coefficients for the energy density,
trace anomaly and entropy, too. The speed of sound receives no improvement
factor at tree level. Note, that these improvement coefficients are exact only
at tree-level, thus in the infinitely high temperature, non-interacting case.
As we decrease the temperature, corrections to these improvement coefficients
appear, which have the form $1+b_2(T)/N_t^2+...$. Empirically one finds that
the $b_2(T)$ coefficient, which describes the size of lattice artefacts of the
tree-level improved quantities, is tiny not only at very high temperatures,
but throughout the deconfined phase. In Section \ref{result} we will present
our findings for different lattice spacings on the effectivity of this
improvement.

\DOUBLEFIGURE{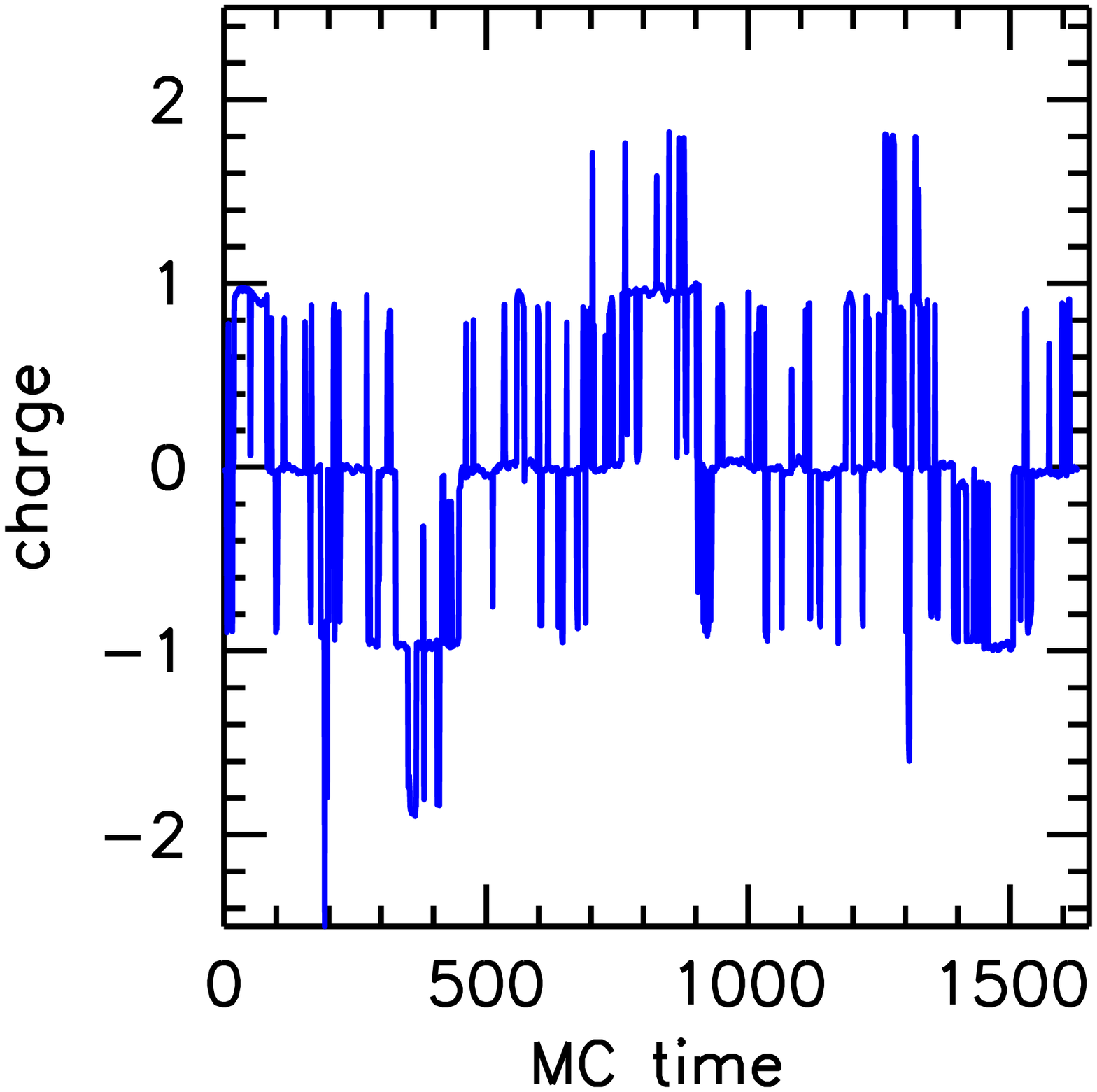,width=7.2cm}{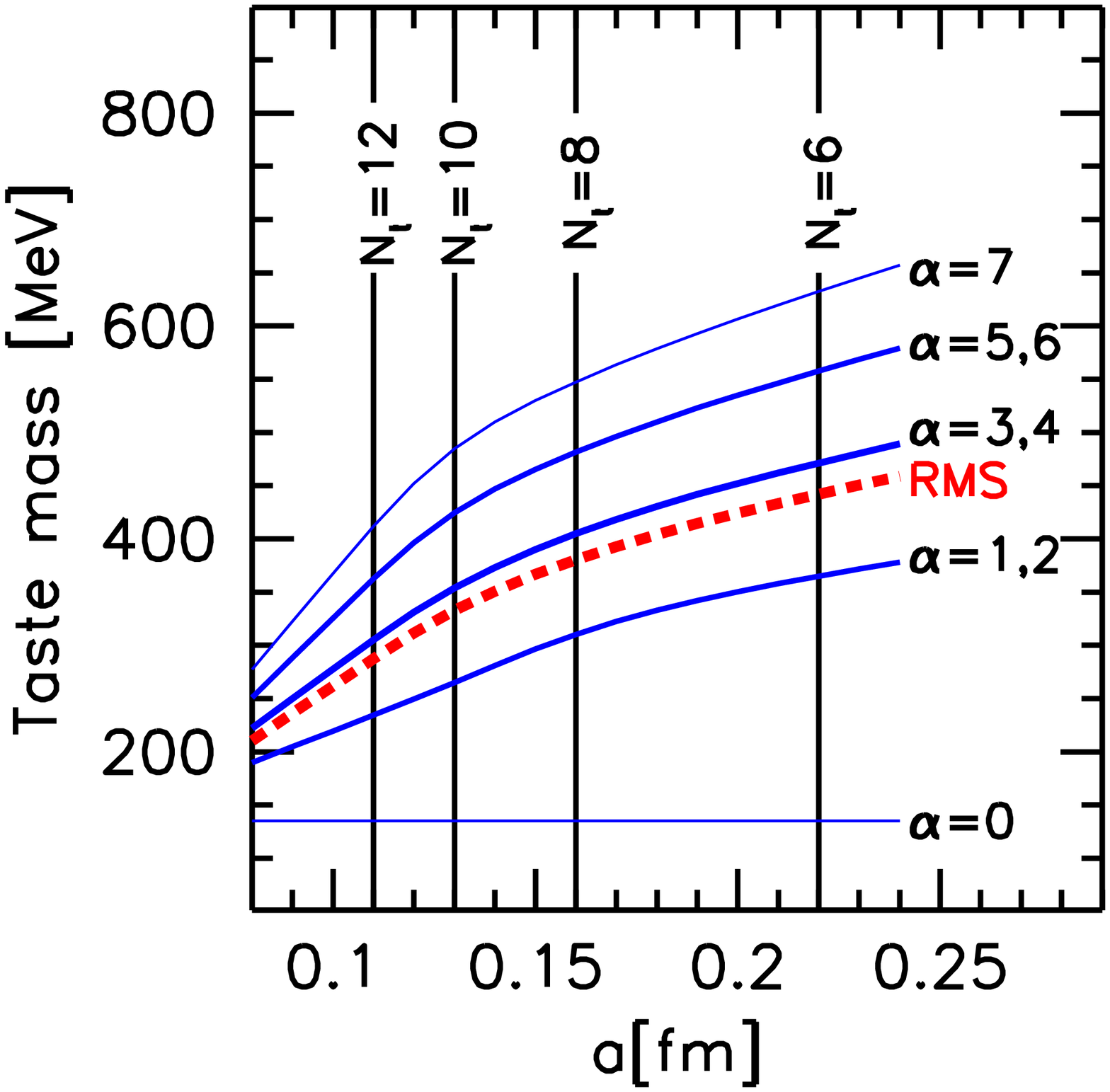,width=7.2cm}
{\label{fig:charge}
Monte-Carlo time history of the topological charge in a zero temperature run
at $a=0.054$ fm lattice spacing.
}
{\label{fig:splitting}
Masses of lattice pion tastes as the function of the lattice spacing. The different $N_t$'s correspond
to $T=150$ MeV.
}
Taste symmetry breaking is a discretization error which is important mainly at
low energies.  In the staggered fermion formulation, hadron masses cannot be
uniquely determined at any finite lattice spacing \cite{Ishizuka:1993mt}. Each
continuum hadron state has a corresponding multiplet of states on the lattice:
due to the taste symmetry violation the masses of these states are split up. As
an example, 16 lattice states correspond to each continuum pion state, each of
them contributing with a 1/16 weight. The following table lists the members of
the lattice pion multiplet with the taste structure (a $4\times4$ complex
matrix, $\Gamma_\alpha$) and the multiplicity ($n_\alpha$):
\begin{equation}
\begin{array}{|c||c|c|c|c|c|c|c|c|}
\hline
\alpha        & 0 & 1 & 2 & 3 & 4 & 5 & 6 & 7 \\
\hline
\Gamma_\alpha & \gamma_5 & \gamma_0\gamma_5 & \gamma_i\gamma_5 & \gamma_i\gamma_j & \gamma_i\gamma_0 & \gamma_i & \gamma_0 & 1 \\
\hline
16\cdot n_\alpha      & 1 & 1 & 3 & 3 & 3 & 3 & 1 & 1 \\
\hline
\end{array}
\nonumber
\end{equation}
Only $\alpha=0$ behaves like a Goldstone-boson, i.e. its mass vanishes in the chiral
limit. The other 15 states have masses of the order of several hundred MeVs for
sensible values of the lattice spacing. Though these mass differences vanish
in the continuum limit, it is very important to suppress them as much as
possible. The effect of the heavier ``pions'' on the equation of state can be
significant: they can reduce the QCD pressure and can also shift the
transition temperature.  Strategies for the suppression have been studied
extensively. In general, using gauge link smearing in the fermion action proved
to be a very efficient and cheap option.  The mass splitting in the pion
multiplet for our smearing recipe as a function of the lattice spacing is
shown on Figure \ref{fig:splitting}. The pion state with the lowest mass is
adjusted to the mass of the continuum pion. We plot the root mean squared (RMS)
average of the masses with red dashed line. The four vertical lines correspond
to lattice spacings on our four different $N_t$'s in the transition region (at
$T=150$ MeV). 

%% file: methods_lcp.tex
\TABLE{
\begin{tabular}{|c|c|c|}
\hline
a\,\mbox{[fm]}&$\beta$&$m_s^{\rm phys}(\beta)$\\
\hline\hline
0.2824(6)&3.45  & $1.57\cdot10^{-1}$\\
0.2173(4)&3.55  & $1.02\cdot10^{-1}$\\
0.1535(3)&3.67  & $6.39\cdot10^{-2}$\\
0.1249(3)&3.75  & $5.03\cdot10^{-2}$\\
0.0989(2)&3.85  & $3.94\cdot10^{-2}$\\
\hline                 
0.0824&3.938(3) & $3.30\cdot10^{-2}$\\
0.0687&4.036(5) & $2.71\cdot10^{-2}$\\
0.0572&4.140(6) & $2.24\cdot10^{-2}$\\
0.0477&4.255(8) & $1.88\cdot10^{-2}$\\
0.0397&4.360(12)& $1.59\cdot10^{-2}$\\
0.0331&4.486(11)& $1.31\cdot10^{-2}$\\
0.0276&4.615(9) & $1.04\cdot10^{-4}$\\
\hline
\end{tabular}
\caption{\label{tab:lcp}
The line of constant physics used in this work. The upper half of this table
quotes the figures of Reference \cite{Aoki:2009sc}, the lower half is the
result of our iterative scheme.  For $a$, $m_s^{\rm phys}$ we estimate the systematic
errors to be 2\%.
}
}
On a lattice with fixed $N_t$ we change the temperature by changing the lattice
spacing. This can be achieved by varying the bare parameters of the action:
$\beta$ and the quark masses. The fact that towards the continuum limit the
lattice should reproduce the continuum physics, dictates the functional
relation between these parameters. Our LCP was defined so, that the mass and
decay constant of the lightest staggered kaon and the mass of the lightest
staggered pion are related to each other as the corresponding continuum
values\footnote{We take $f_K=155.5$ MeV, $m_\pi=135$ MeV and $m_K=495$ MeV from
the Particle Data Group \cite{Amsler:2008zzb}.}. This then translated to a
quark mass ratio $m_s/m_{ud}=28.15(1)$ and to the functions $m_s^{\rm
phys}(\beta)$ and $a(\beta)$ (see upper part of Table \ref{tab:lcp}).  When we
say that quark masses are set to their physical values, we mean that they are
on this LCP ($m_{ud}^{\rm phys}(\beta)$ and $m_s^{\rm phys}(\beta)$). For
details see References \cite{Aoki:2009sc,Aoki:2006br}, where the LCP was
determined in the range $\beta=3.45\dots3.85$. In the following, we refer to
this technique as our ``old method''.

In this work we determine the LCP up to $\beta\approx 4.62$, which allows us to
calculate thermodynamical observables up to temperatures as high as $\sim 1$
GeV.  Due to the smallness of the lattice spacing at these $\beta$ values, the
calculation of hadronic observables is impossible with the present computer
resources. Thus the conventional way (our ``old method'') to determine the LCP by measuring ratios
of hadronic observables fails here. We circumvented the problem by applying an
iterative procedure analogous to the well-known step scaling technique
\cite{Luscher:1991wu}. In order to simplify the discussion, first we introduce the
technique without quarks, ie. in the pure gauge case. The inclusion of quarks
will be described afterwards. Later on we refer to this technique as our ``new method''.

We make simulations on symmetric lattices ($N^4$) with $N\le24$. For small
lattice spacings they are in the ``deconfined phase'', no hadronic observable
can be used for scale setting here. Therefore we use the Creutz ratio ($\chi$)
for the scale setting similarly to what was proposed in \cite{Bilgici:2009kh}.
On a $N^4$ lattice at coupling $\beta$ we define an effective coupling by
measuring symmetric Creutz ratios\footnote{We use APE smearing to enhance the
signal/noise ratio \cite{Albanese:1987ds}.} of size $N/4$:
\be
\label{eq:creutzgsqr}
g^2_{\rm eff}(N,\beta) = N^2 \chi(N/4). 
\ee
Let us assume, that we have already determined the LCP, in this case a single
$\beta(a)$ function, for lattice spacings larger than $a_0$. This is obtained
by fixing some observables in the hadronic phase (e.g. $r_0$ or glueball mass).
Now we want to determine this function for a lattice spacing smaller than
$a_0$. We take lattices of size $12^4$, $16^4$ and $20^4$ with different
lattice spacings, so that the size of the lattices in physical units is to be
the same, which is at couplings $\beta(20/12\cdot a_0)$, $\beta(20/16\cdot
a_0)$, $\beta(20/20\cdot a_0)$. Then we make a continuum extrapolation from the
measured effective couplings to $a_1= 20/24\cdot a_0 < a_0$ and search for the
$\beta$ parameter at which the $g^2_{\rm eff}$ is reproduced on a $24^4$
lattice.  This defines $\beta(a_1)$ and completes one step of iteration.  The
next iteration repeats the above steps using $a_1$ instead of $a_0$. Thus, one
receives a series of $\beta$ values, which correspond to smaller and smaller
lattice spacings: $a_{i+1}=20/24\cdot a_i$.

As a test, we used this method to determine the $\beta(a)$ function in the pure
gauge case using the plaquette gauge action. We initialized the first iteration
with $\beta<6.0$ data. For comparison, we take the simulations up to
$\beta<6.92$ from \cite{Necco:2001xg} and an extrapolation based on the results
of \cite{Gockeler:2005rv}. The agreement is maintained over more than an order
of magnitude in the lattice spacing (see Figure \ref{fig:a_que}).
\DOUBLEFIGURE{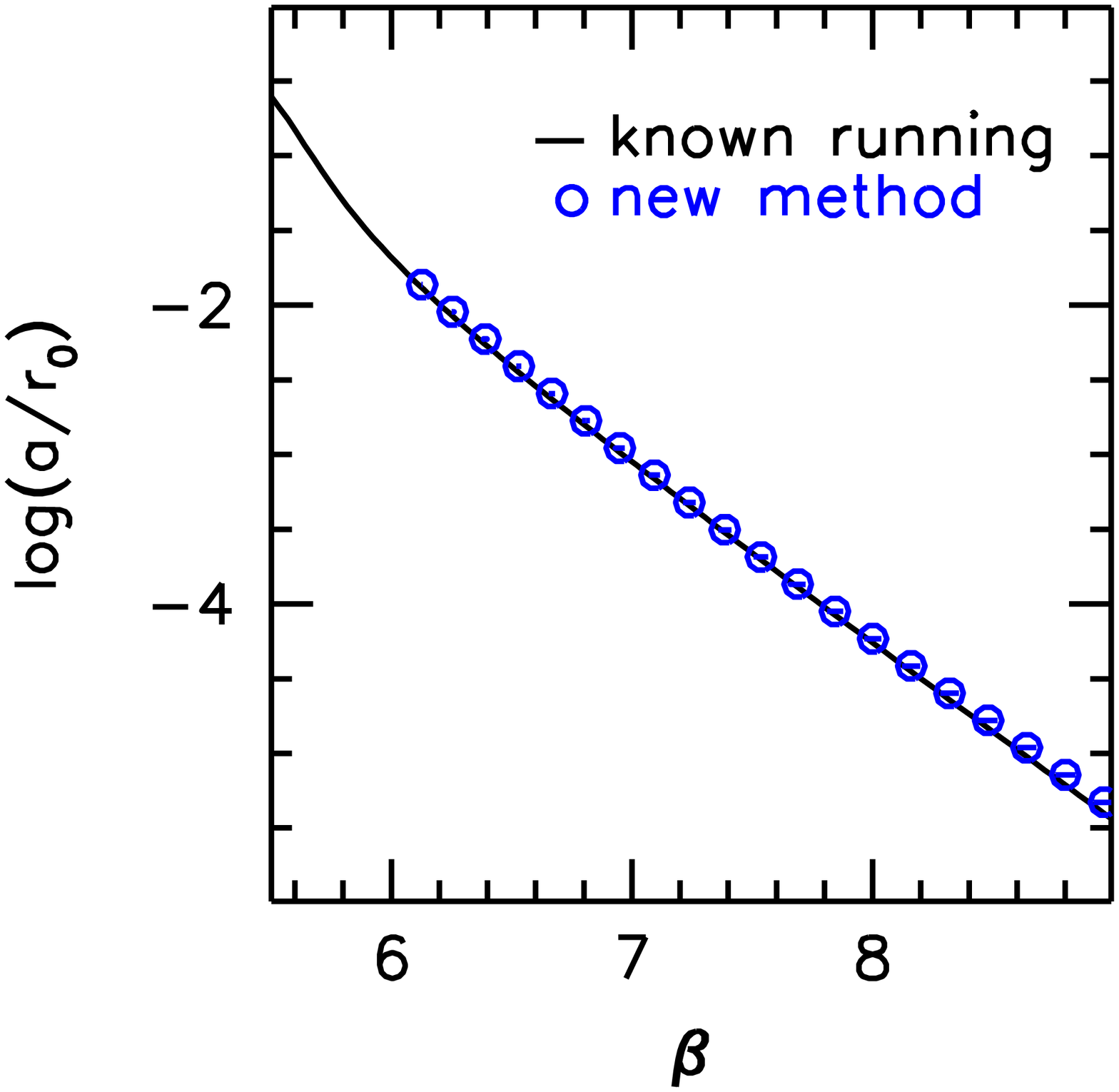,width=7.2cm}{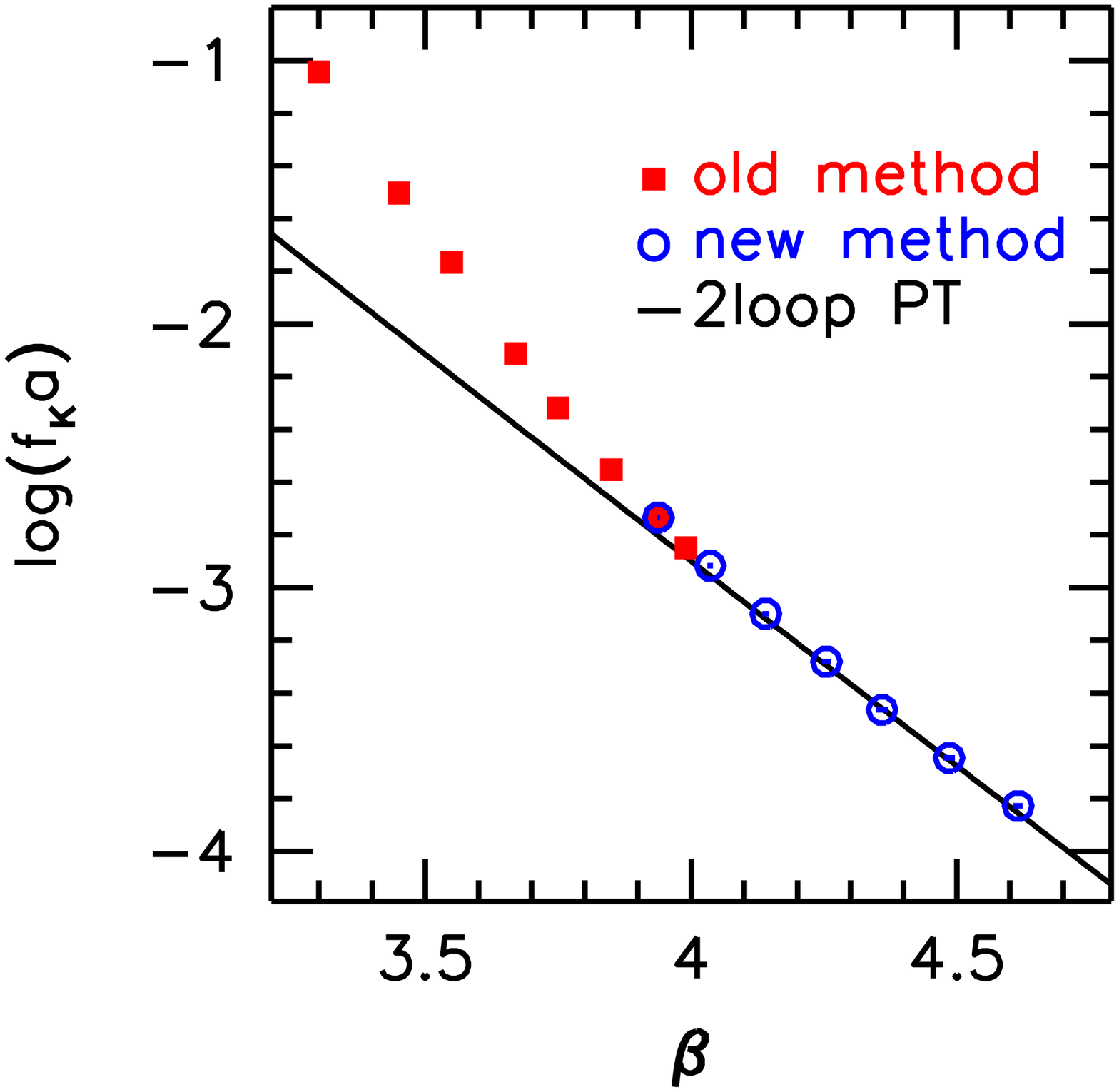,width=7.2cm}
{\label{fig:a_que}
Lattice spacing as a function of $\beta$ in the pure gauge case using the Wilson action. The
points are obtained with our variant of the step scaling method, the line is the already known running.
}
{\label{fig:a_dyn}
Lattice spacing as a function of $\beta$ in the dynamical $n_f=2+1$ case for our action. The blue circles are obtained with our new
method, the red squares with the old one by measuring hadronic observables. The black curve is the 2-loop perturbative running.
}

If one would like to include the quarks properly, a renormalized observable,
that is not related to the hadronic scale and is sensitive to the quark masses
would be necessary. The above iterative procedure then would have to be
generalized, where not only the $\beta$ but the quark masses would have to be
tuned. Since we know, that for small enough lattice spacings the running of the
scale does not depend on the quark masses and that for high temperatures the
thermodynamics shows a very small quark mass dependence, the quark masses can
be determined with less precision. Therefore from $\beta>3.94$ we have
abandoned this multi-parameter tuning and simply used the 1-loop running of the
quark mass in the simulations\footnote{The running of the quark mass in the
range $\beta=3.85\dots3.94$ was determined by tuning the $\beta$ and the quark
masses and monitoring the pseudoscalar meson masses.}. The obtained results are
tabulated in the lower part of Table \ref{tab:lcp}. Since the new method
measures the $\beta$ at a given lattice spacing, the errors are attributed to
the $\beta$ values. The running of the coupling is shown on Figure
\ref{fig:a_dyn}. Note, that we have an overlap between the lattice spacings
obtained by the ``old'' and the ``new'' methods.  Together with the simulation points
with these techniques, we also show the two-loop perturbative running. A
satisfactory agreement is found for $\beta>4.1$.

At $\beta=3.99$ we have checked the resulting LCP using a zero temperature
simulation with physical quarks on a $64^4$ lattice. On Figure \ref{fig:a_dyn}
the highest $\beta$ point with the old method corresponds to this simulation.
Setting the scale by $f_K$ we get results compatible with the physical hadron
spectrum:
\begin{equation}
\begin{array}{|c|c|c|c|}
\hline
m_\pi & m_K & m_{\eta_s} & f_\pi \\
\hline
134.2(6) {\rm ~MeV} & 497.7(6) {\rm ~MeV} & 690(4) {\rm ~MeV} & 131.0(7) {\rm ~MeV} \\
\hline
\end{array}
\nonumber
\end{equation}
whereas for the lattice spacing we find $a=0.0735(4)$ fm. These results justify
the correctness of our new technique to determine the LCP. In the range
$\beta= 3.85\dots4.1$ we also performed control simulations in smaller volumes
(with $N\approx 2$ fm). These showed up to 2\% deviations from the line of
constant physics determined with the new method.  This gives an estimate of the
systematic errors coming from scale setting used in this work.

We also study the thermodynamics of QCD for heavier than physical quark masses.
The corresponding renormalized trajectories are defined by setting the strange
mass to the above obtained $m_s^{\rm phys}(\beta)$, whereas the light quark
mass is set to $m_{ud}(\beta)= m_s^{\rm phys}(\beta)/R$, where the quark mass
ratio $R$ is varied in the range $R=1\dots 28.15$.  Since these are
hypothetical theories with nonphysical quark masses, strictly speaking they are
not related to scales in physical units (e.g. fm). Nevertheless, it is useful to
use physical scales also in this case. To this end, for heavier than physical
quark masses we use the standard mass independent scale setting, thus
$\beta(a)$ is the same as for physical quark masses.

Furthermore, we also give an estimate for the contribution of the charm quark to
thermodynamics. The charm quark mass is not very well known experimentally,
whereas there is a recent high precision lattice calculation of the charm to
strange mass ratio \cite{Davies:2009ih}. We denote this ratio by $Q=m_c/m_s$
and the recent lattice estimate is $Q=11.85(16)$. In addition to the central
value of this lattice result, we carried out the analysis for several other
values in the range $Q=10.75\ldots 20$. The effect of the charm was determined
at the partially quenched level, ie. its back-effects on the gauge background
were neglected. We leave it for future work to include the charm quark
dynamically.

%% file: methods_int.tex
In order to fix the notations we first quickly review the standard method, which is
used to determine the equation of state (``integral technique'', \cite{Engels:1990vr}). On the
lattice, the dimensionless pressure
\be
p^{\rm lat}(\beta,m_q)=(N_tN_s^3)^{-1}\log\mathcal{Z}(\beta,m_q)
\ee
is in itself not accessible using conventional algorithms, only its derivatives
with respect to the bare parameters of the action are measurable. Using these
partial derivatives the pressure can be rewritten as a
multidimensional integral along a path in the space of bare parameters:
\be
p^{\rm lat}(\beta,m_q)-p^{\rm lat}(\beta_0,m_{q0})=
(N_tN_s^3)^{-1} \int^{(\beta,m_q)}_{(\beta_0,m_{q0})}
\left(
d\beta \frac{\partial \log \mathcal{Z}}{\partial \beta} +
\sum_q d m_q \frac{\partial \log \mathcal{Z}}{\partial m_q}\right),
\ee
where we used the index `0' to denote the starting point of the integration.
The value of the pressure at parameters `0' has to be handled with
care: one either chooses the starting point so deeply in the strong coupling
regime, so that the $p^{\rm lat}(\beta_0,m_{q0})$ can already be neglected or
it can be taken from model computations (see Subsection
\ref{methods_hrg}). The derivatives in this formula are the gauge action and
the chiral condensate densities:
\be
\langle -s_g\rangle= (N_tN_s^3)^{-1}\frac{\partial \log\mathcal{Z}}{\partial \beta},
\quad \quad \langle\bar\psi_q\psi_q\rangle=(N_tN_s^3)^{-1}\frac{\partial \log\mathcal{Z}}{\partial m_q}.
\label{eq:ders}
\ee

The pressure itself contains additive divergences, which are independent of the
temperature. Removing them can be done by subtracting the same observables
measured on a lattice, with the same bare parameters but at a different temperature
value, ie. with a different temporal extent $N_t^{\rm sub}$. Here we use lattices with
a large enough temporal extent, so it can be regarded to have zero temperature.
These are denoted as ``zero $T$'' lattices in Subsection \ref{methods_act}. The
finiteness of $N_t^{\rm sub}$ causes an error of the order $(N_t/N_t^{\rm
sub})^4$, which is smaller than the typical size of the statistical error.
Using the subtracted observables 
\begin{eqnarray}
\langle s_g\rangle^{\rm sub}= \langle s_g\rangle_{N_t,N_s} - \langle s_g\rangle_{N_t^{\rm sub},N_s} \\
\langle\bar\psi_q\psi_q\rangle^{\rm sub}= 
\langle\bar\psi_q\psi_q\rangle_{N_t, N_s} - \langle\bar\psi_q\psi_q\rangle_{N_t^{\rm sub}, N_s}
\end{eqnarray}
one can express the renormalized pressure as:
\be
\label{eq:intp}
\frac{p(T)}{T^4}-\frac{p(T_0)}{T_0^4}=
N_t^4 \int^{(\beta,m_q)}_{(\beta_0,m_{q0})}
\left(d\beta \langle -s_g \rangle^{\rm sub} +
\sum_q d m_q \langle \bar\psi_q\psi_q\rangle^{\rm sub}\right),
\ee
where we have normalized it by $T^4$ as usual. The $T$ and $T_0$ are the
temperature values corresponding to the lattice spacing at the bare parameters
$(\beta,m_q)$ and $(\beta_0,m_{q0})$. Using Equation (\ref{eq:Idef})
one can also relate the integrand to the trace anomaly:
\be
\frac{I(T)}{T^4} \frac{dT}{T}=
N_t^4\left(d\beta \langle -s_g \rangle^{\rm sub} +
\sum_q d m_q \langle \bar\psi_q\psi_q\rangle^{\rm sub}\right).
\label{eq:tracealat}
\ee
Let us make an observation here: due to the $N_t^4$ prefactor in the
trace anomaly the subtracted observables decrease with the lattice
spacing (at a fixed temperature this means an increasing $N_t$). While
$\langle -s_g \rangle^{\rm sub}$ decreases as $N_t^{-4}$, the chiral condensate
behaves substantially different. Due to chiral symmetry\footnote{Staggered
fermions have only a remnant chiral symmetry, but this does not affect the
argument.} $\langle \bar\psi_q\psi_q \rangle$ is proportional to the bare quark
mass, moreover it is also multiplied by another power of the bare quark mass
($dm_q$ line-element) in the trace anomaly. Since the bare mass also
decreases with the lattice spacing, the subtracted condensate
only decreases as $N_t^{-2}$. These scalings are directly translated into different precisions
for the two terms in Equation (\ref{eq:tracealat}). The term with the gauge
action density has $N_t^2$ times larger errors than the term with the chiral
condensate, if they are evaluated with the same statistics. 

The standard integral method proceeds as follows: first one calculates the
trace anomaly for several temperatures along the lines of constant physics and
then integrates it to get the pressure up to an integration constant (which can
be either neglected or taken from a model calculation).  This path was used in
several lattice studies, e.g.
\cite{Cheng:2007jq,Bazavov:2009zn,Cheng:2009zi,Aoki:2005vt}.

One of our major goals with this paper is to determine the equation of state for
several (heavier than physical) different values of the light quark masses.  We
therefore carried out simulations covering an extended region in the
$\{\beta,m_s,m_{ud}\}$ parameter space. The strange mass $m_s$ was always set to its physical
value $m_s^{\rm phys}(\beta)$ and $m_{ud}$ was set as
$m_{ud}(\beta)=m_s(\beta)/R$ with the quark mass ratio $R$ ranging from $1$ to
$R^{\rm phys}=28.15$. The value $R=1$ corresponds to the three degenerate
flavor case, whereas $R^{\rm phys}$ is the ``real world'' value of the quark
mass ratio.  At these parameters we can consider the pressure as a function of
only two variables: $\beta$ and $R$. The respective derivatives can be measured
on the lattice, they are given by the following formulas:
\begin{eqnarray}
\label{eq:dbeta}
D_\beta= \left.\frac{\partial }{\partial \beta}\right|_{R}\frac{p(T)}{T^4}&=& N_t^4 \left[
\langle -s_g \rangle^{\rm sub} + 
\frac{\partial m_s^{\rm phys}}{\partial \beta}(\langle \bar\psi_{s}\psi_{s}\rangle^{\rm sub} +
\frac{1}{R}\langle \bar\psi_{ud}\psi_{ud}\rangle^{\rm sub})\right],\\
\label{eq:dR}
D_R= \left.\frac{\partial }{\partial R}\right|_{\beta}\frac{p(T)}{T^4}&=& -N_t^4
\frac{m_{s}^{\rm phys}}{R^2}\langle \bar\psi_{ud}\psi_{ud}\rangle^{\rm sub}.
\end{eqnarray}
These derivatives have to be integrated to obtain the pressure. Since the
integrand is in itself a total derivative, the result is independent of the
chosen integration path in the two dimensional space of the parameters $\beta$ and $R$
(see Figure \ref{fig:int1} for an illustration). The simulation
points are denoted by crosses. The straightforward way would be to perform a
one dimensional integral in $\beta$ at a fixed value of the quark mass ratio
$R$, this corresponds to path A.  However one can also imagine zigzagging
routes, like path B or C on the Figure. Averaging several of such integrals,
one can increase statistics and it also provides an estimate of the systematic
error related to the integration path.

\EPSFIGURE{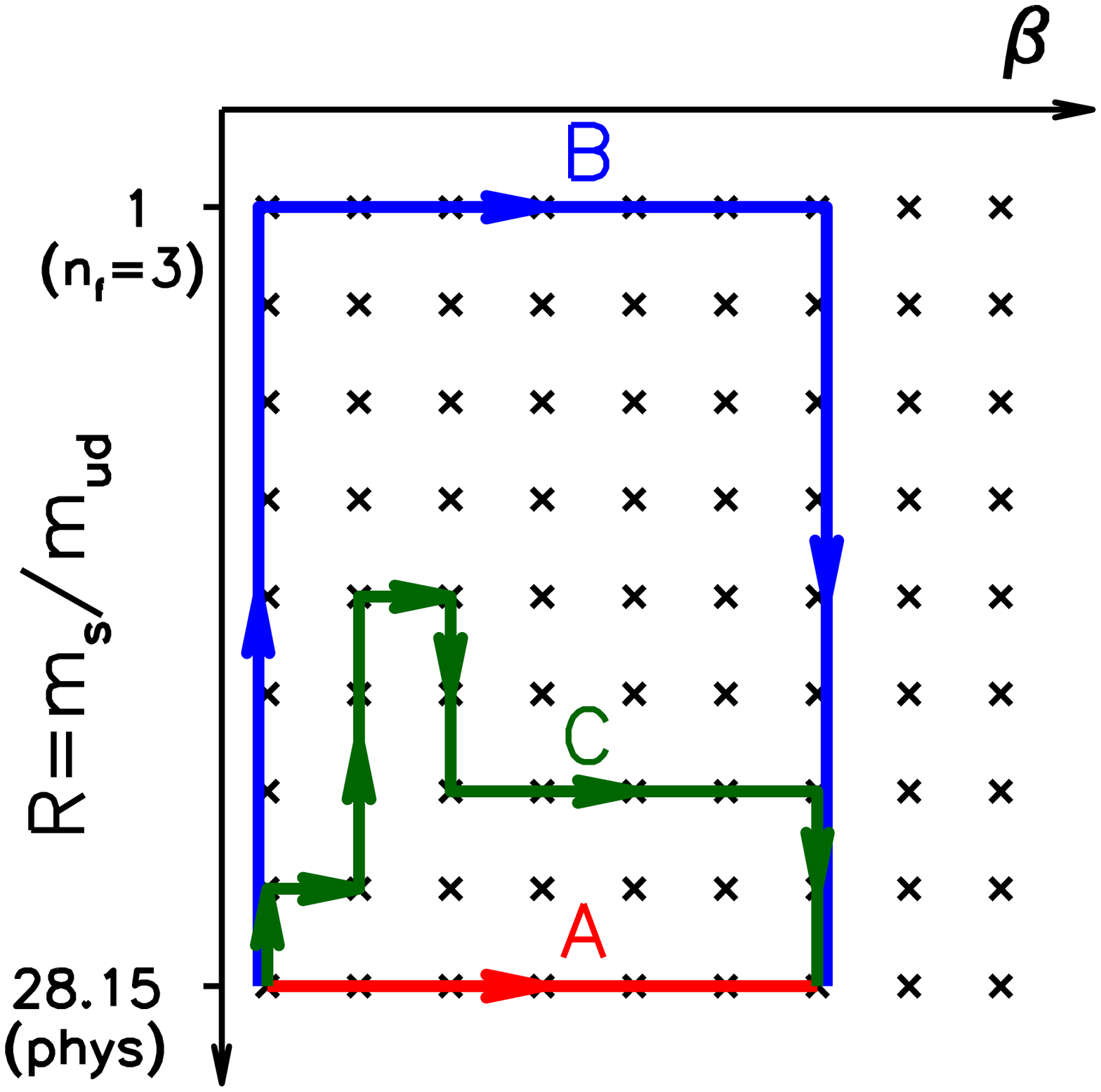,width=7.0cm}
{\label{fig:int1}
Illustration of possible integration paths, which can be used to obtain the
pressure at a certain point. For explanation see the text.
}
Here we propose a generalized method that takes into account every possible
integration path at the same time. The main idea is -- instead of parametrizing
the derivatives of the pressure and then integrate -- to parametrize the
pressure itself. A straightforward parametrization is one using the actual
values $p_{kl}$ of the pressure at some node points $\{\beta_k,R_l\}$ that
build up a two-dimensional spline function -- i.e. that unambiguously determine
the whole pressure surface as a function of $\beta$ and $R$. The parameters
$p_{kl}$ can be set to minimize the deviation between the derivatives of this
surface and the measured values $D_\beta$ and $D_R$. This minimum condition
leads to a system of linear equations that can be solved for $p_{kl}$. This
method gives the pressure directly using the information contained in the
derivatives $D_\beta$ and $D_R$ without the need to carry out an actual
integration. The details of this analysis can be found in Appendix
\ref{app_spline} and in Reference \cite{Gergospline}, where the systematic
error of the method is also discussed.

The method determines the pressure only up to an overall constant factor,
which corresponds to an integration constant and originates from the fact that
the derivatives $D_\beta$ and $D_R$ are measured at finite values of the
temperature. We set this constant, so that at the smallest temperature for
three degenerate flavors the pressure is set to zero:
\be
\label{eq:nullpt}
p(T=100 {\rm MeV}, R=1)= 0.
\ee
The error of this choice is hard to estimate from the lattice alone. It is
definitely more reasonable to set the zero here, than at the physical point
$R=R^{\rm phys}$, since the hadrons are much heavier here and therefore their
contribution to the pressure is smaller. According to the HRG
model (see Subsection \ref{methods_hrg}) the pressure at this point 
$p^{\rm HRG}(T)/T^4= 0.02$ is
much smaller than the typical statistical errors on the lattice.

Starting from the above obtained parametrization of the pressure we derived the
trace anomaly using Equation (\ref{eq:Idef}). In order to be able to give
smooth curves as results, we fitted the trace anomaly with a 4-degree spline
function. In this way the smoothness of the further derived quantities
($\epsilon$, $s$ and $c_s$) was ensured, too. The variation of the nodepoints
of this spline interpolation was used to estimate the systematic error. 

The above described method was used to determine the equation of state on
$N_t=6,8$ lattices for different values of the quark masses. In the case of
$N_t=10$ we determined the equation of state exclusively with physical quark
masses. In order to satisfy the normalization condition in Equation
(\ref{eq:nullpt}), we made heavier mass simulations at $T=100$ MeV up to the three
flavor point. On $N_t=12$ we made simulations at three temperature
values, this allows us to calculate the trace anomaly only.

For $N_t=8$ we also estimated the size of the contribution of the charm
quark. In order to obtain this contribution, the $\beta$ derivative of the
pressure in Equation (\ref{eq:dbeta}) has to be modified by adding the charm
condensate:
\be
D_{\beta} \to D_{\beta} + \frac{\partial m_s}{\partial \beta}\cdot Q \cdot \langle \bar\psi_c \psi_c \rangle^{\rm sub}.
\ee
Let us emphasize it here again, that we use the partially quenched
approximation, when calculating the observables in the $n_f=2+1+1$ theory: we
neglect the back effect of charm quarks on the gauge fields.

%% file: methods_hrg.tex
The Hadron Resonance Gas model has been widely used to study the hadronic
phase of QCD in comparison with lattice data
\cite{Karsch:2003vd,Karsch:2003zq,Tawfik:2004sw}.  The low temperature phase is dominated by
pions. Goldstone's theorem implies weak interactions between pions at low
energies, which allows to study them within Chiral Perturbation Theory
($\chi$PT) \cite{Gasser:1983yg}. As the temperature $T$ increases, heavier states
become more relevant and need to be taken into account. The HRG
model has its roots in the theorem by Dashen, Ma and Bernstein
\cite{Dashen:1969ep}, which allows to calculate the microcanonical partition
function of an interacting system, in the thermodynamic limit
$V\rightarrow\infty$, to a good approximation, assuming that it is a gas of
non-interacting free hadrons and resonances \cite{Venugopalan:1992hy}. 
 
The pressure of the HRG model can be written as the sum of
independent contributions coming from non-interacting resonances
\begin{eqnarray}
\frac{p^{\rm HRG}}{T^4}
=\frac{1}{VT^3}\sum_{i\in\;mesons}\hspace{-3mm} 
\log{\mathcal{Z}}^{M}(T,V,\mu_{X^a},m_i)
+\frac{1}{VT^3}
\sum_{i\in\;baryons}\hspace{-3mm} \log{\mathcal{Z}}^{B}(T,V,\mu_{X^a},m_i)\; ,
\label{eq:ZHRG}
\end{eqnarray}
where
\begin{eqnarray}
\log{\mathcal{Z}}^{M}(T,V,\mu_{X^a},m_i)
&=&- \frac{V {d_i}}{{2\pi^2}} \int_0^\infty dk k^2
\log(1- z_ie^{- \varepsilon_i/T}),
\nonumber\\
 \log{\mathcal{Z}}^{B}(T,V,\mu_{X^a},m_i)
 &=&\frac{V {d_i}}{{2\pi^2}} \int_0^\infty dk\, k^2
\log(1+ z_ie^{- \varepsilon_i/T})  ,
\label{eq:ZMB}
\end{eqnarray}
with energies
$\varepsilon_i=\sqrt{k^2+m_i^2}$, degeneracy 
factors $d_i$ and fugacities
\begin{eqnarray}
z_i=\exp\left((\sum_a X_i^a\mu_{X^a})/T\right) \; .
\label{eq:fuga}
\end{eqnarray}

In the above equation, $X^a$ are all possible conserved charges, including the
baryon number $B$, electric charge $Q$ and strangeness $S$. The sums in
Equation (\ref{eq:ZHRG}) include all known baryons and mesons up to 2.5 GeV, as
listed in the latest edition of the Particle Data Book \cite{Amsler:2008zzb}.
Notice, that only a few resonant states with masses larger than $2$ GeV have
been identified and got into the listing of the Particle Data Group. Attempts
to improve the HRG model by including an exponential mass spectrum for these
heavy resonances have been proposed
\cite{NoronhaHostler:2007jf,NoronhaHostler:2009cf,Majumder:2010ik}. In the
present analysis we will consider only the known states and not include this
exponential spectrum.

For temperatures $T\lsim$ 60-100 MeV the contribution of the pions dominate the
pressure, for larger temperatures, the kaon contribution becomes also sizable.
Heavier states become relevant for $T\gsim120$ MeV.

\EPSFIGURE{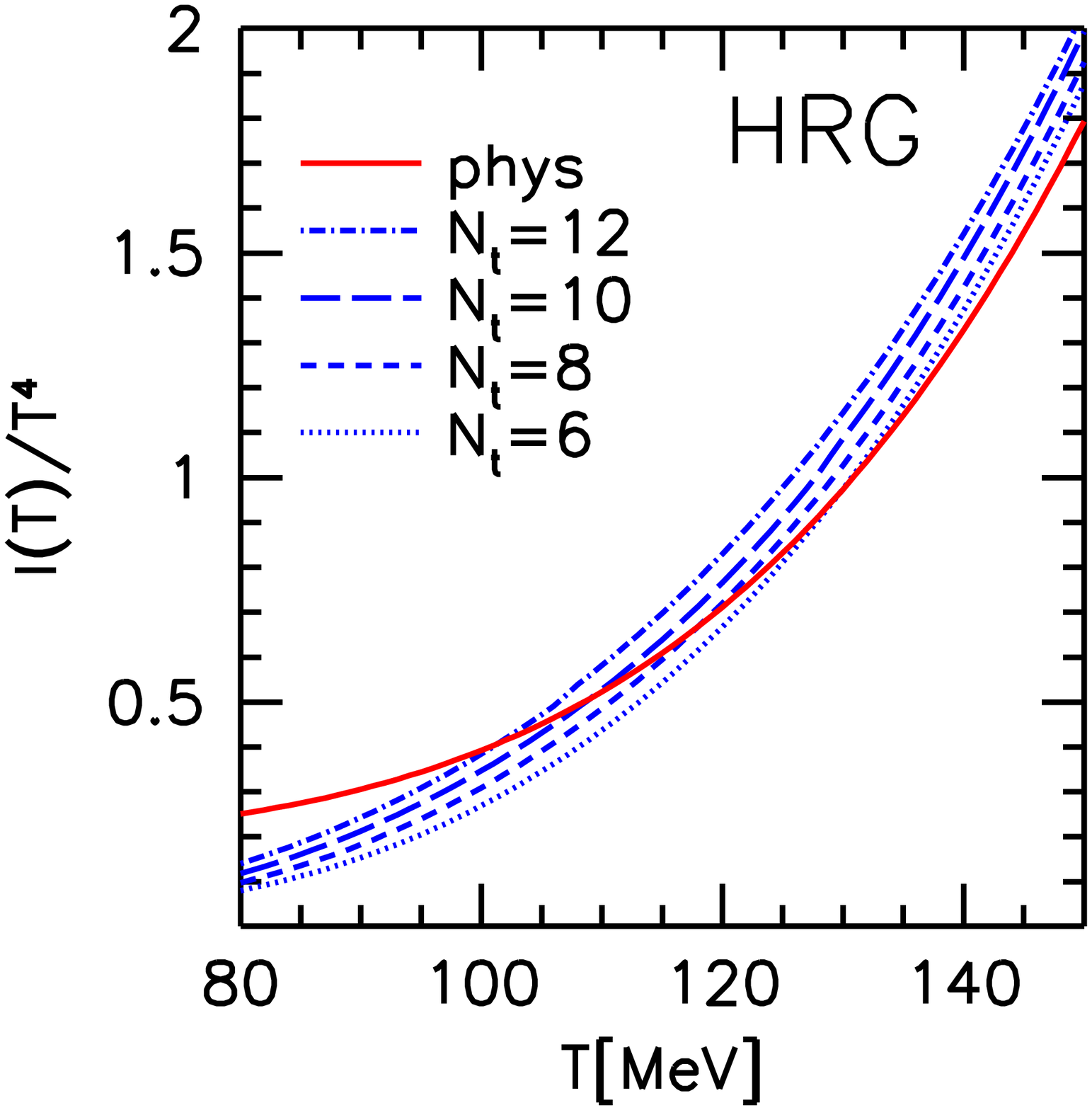,width=7.2cm}
{\label{fig:hrgI}
The normalized trace anomaly in the physical HRG model (solid red line) and 
in the HRG models with the lattice hadron spectrum (dashed blue lines).
}
As we have already discussed in Subsection \ref{methods_act} the staggered
lattice discretization has a considerable impact on the hadron spectrum. In
order to investigate these errors, we define a ``lattice HRG'' model, where in
the hadron masses lattice discretization effects are taken into account.  We
consider only the taste splitting effects for the pions and the kaons. For
example the contribution of the pions to the normalized pressure is now given
by a sum, which runs over the different pion tastes:
\begin{equation}
\frac{1}{VT^3}\sum_{\alpha=0}^{7} n_\alpha
\log{\mathcal{Z}}^{M}(T,V,\mu_{X^a},m_\alpha),
\label{eq:ps}
\end{equation}
where the $n_\alpha$ multiplicities are taken from the table in Subsection
\ref{methods_act}, the $m_\alpha$ masses are those of the different pion tastes
taken from the lattice simulations.

On Figure \ref{fig:hrgI} we plot the normalized trace anomaly of the HRG model
with the physical and with the lattice distorted spectrum for the four
different lattice temporal extensions used in our investigations.  The
difference between the physical and lattice curves is a first estimate of the
lattice discretization errors arising from the taste violation. As the
temperature decreases at a fixed $N_t$, the lattice spacing gets larger and so
do the taste violation effects. The model calculation suggests, that the lattice
results may have sizeable systematic errors in the low temperature region.
Above $T\sim 100$ MeV, this error estimate for the interaction measure is
smaller than the typical magnitude of other errors in lattice QCD calculations.
The overall uncertainties related to the above phenomenon will be discussed
later.

In order to compare the HRG model results with our additional lattice
simulations at larger-than-physical quark masses, we need the pion mass
dependence of all hadrons and resonances included in the calculation. As we
already did in \cite{Borsanyi:2010bp}, we assume that all resonances behave as
their fundamental state hadrons as functions of the pion mass. For the
fundamental hadrons, we use the pion mass dependence from Reference
\cite{Huovinen:2009yb}. For larger-than-physical quark masses, the taste
symmetry violation at finite lattice spacing has a milder impact on the
pressure. This also motivates to take $R=1$ as the starting point of the
integration of the pressure at $T=100$ MeV (see Subsection \ref{methods_int}).

%% file: result.tex
In this section we present our results on the equation of state. First we
discuss finite volume effects and discretization artefacts.  Then we show the
$n_f=2+1$ flavor pressure, the interaction measure, the energy and entropy
density as well as the speed of sound. We identify characteristic points in the
temperature dependence of these observables and we also provide a
parametrization. Moreover, we discuss the mass dependence of the equation of
state. Afterwards we determine the size of the contribution of the charm quark
within the partially quenched framework.  Finally we compare our results to the
existing literature.

For high temperatures the thermodynamic quantities approach the values of the
non-interacting massless relativistic gas, the so-called Stefan-Boltzmann
limit. The limit for the three flavor pressure is $p_{SB}/T^4\approx 5.209$,
for the energy density $\epsilon_{SB}=3p_{SB}$ and for the entropy density
$s_{SB}=4p_{SB}/T$.

The error bars on the figures are obtained by quadratically adding the
statistical error and the systematic error of the integral method (Subsection
\ref{methods_int}). If no error bar is shown on a point, then the errorbar is
smaller than the pointsize. The temperature values have an error at the $2\%$
level arising from the scale setting procedure (Subsection \ref{methods_lcp}). 

\subsection*{Finite volume and discretization effects}
\DOUBLEFIGURE{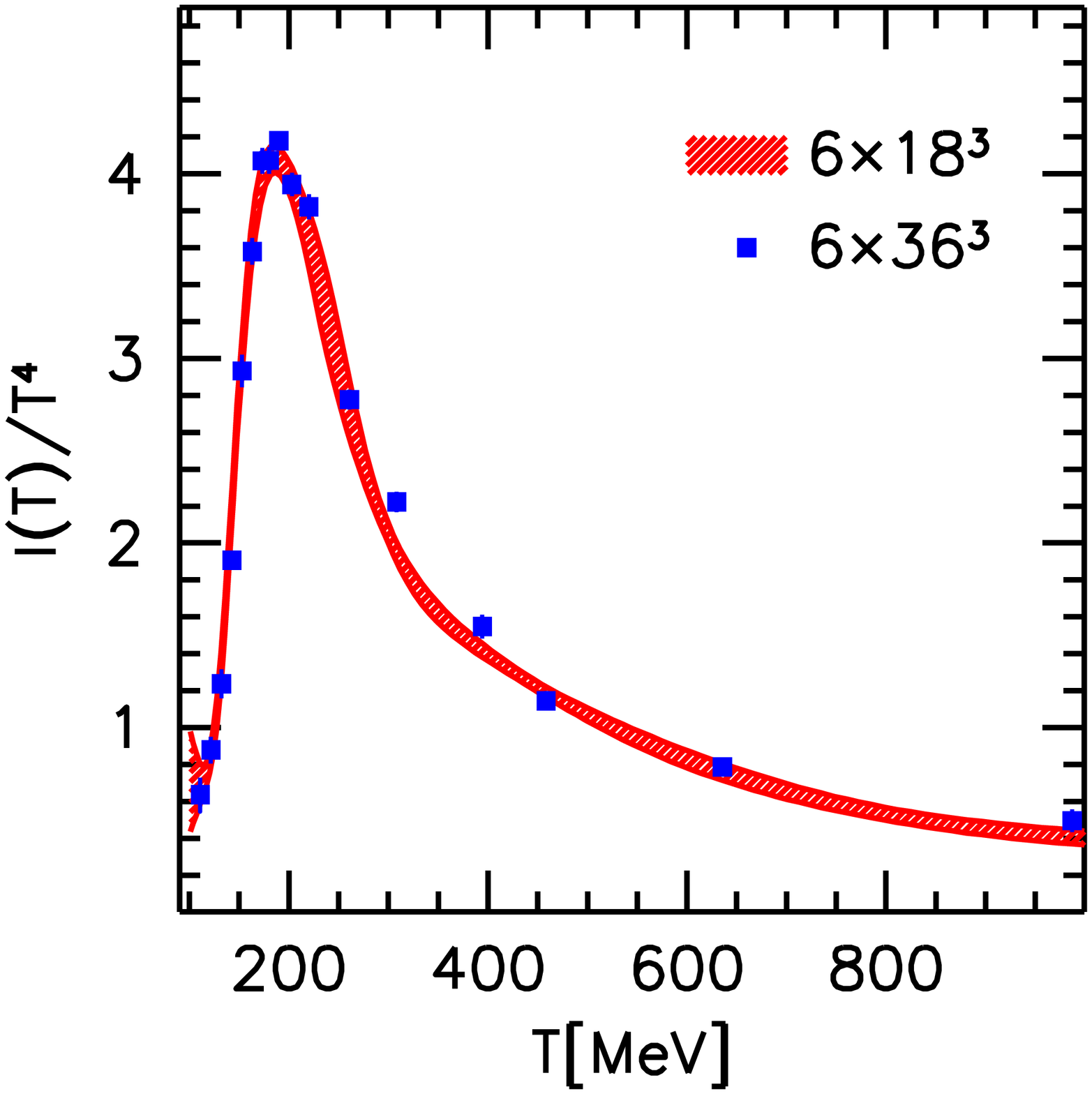,width=7.2cm}{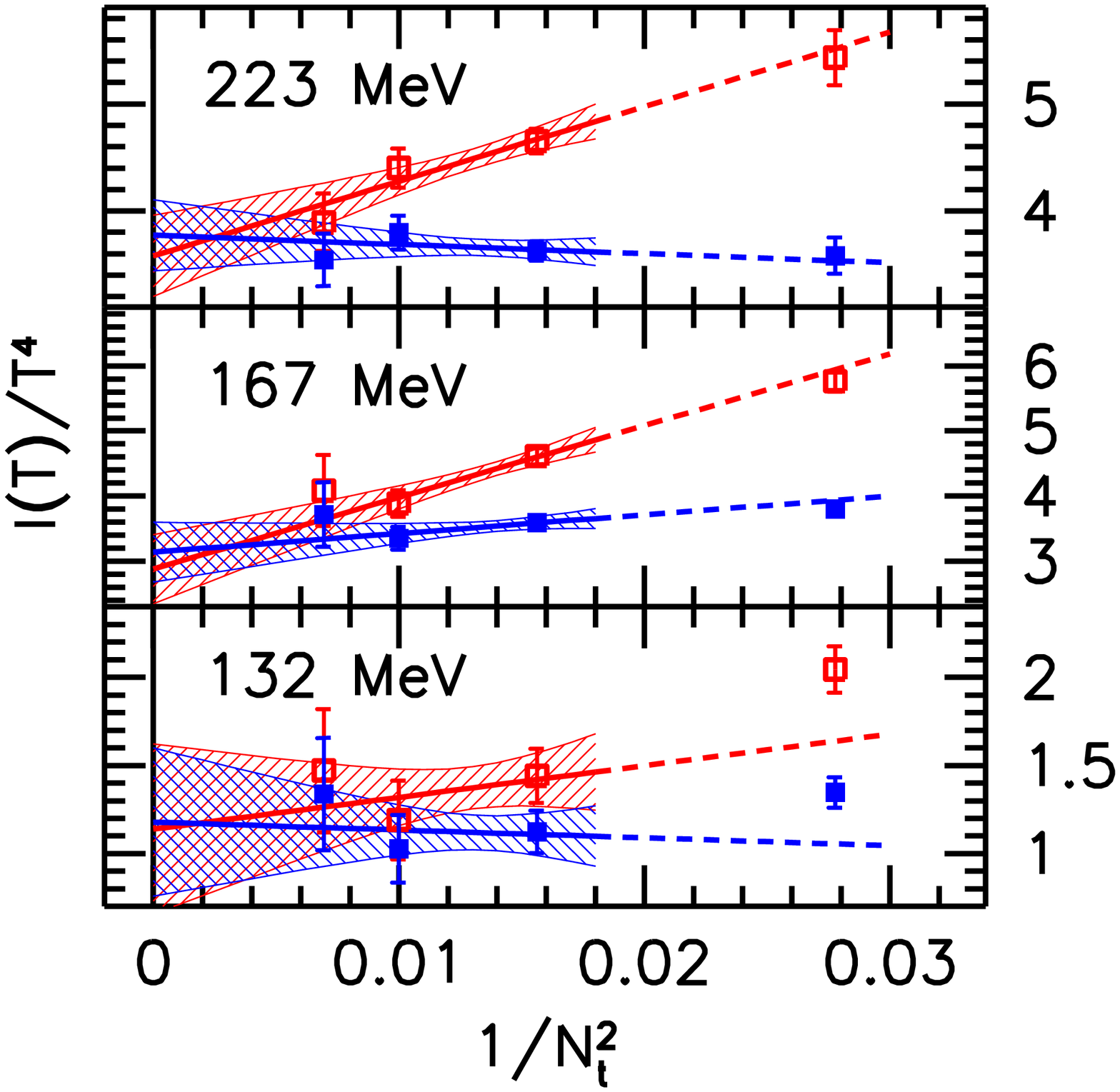,width=7.2cm}
{\label{fig:finV}
The trace anomaly on lattices with different spatial volumes:
$N_s/N_t=3$ (red band) and $N_s/N_t=6$ (blue points). 
}
{\label{fig:rescale}
The trace anomaly at three different temperatures as a
function of $1/N_t^2$. 
Filled blue symbols represent the results within the lattice tree-level improvement
framework, red opened symbols show the results without this improvement. The error
of the continuum extrapolated value is about $0.4$ for all three temperatures.
}

In order to verify that there are no significant finite size effects present in
the lattice data with the aspect ratio $N_s/N_t=3$, we checked our $N_t=6$ data
against a set of high precision $N_s/N_t=6$ simulations. The latter lattice
geometry corresponds to about 7~fm box size at the transition temperature.
Figure \ref{fig:finV} shows the comparison between the two volumes for the
normalized trace anomaly $I/T^4$. From this result we concluded that it is
acceptable to perform the more expensive simulations throughout with
$N_s/N_t=3$. Note, however, that for small temperatures this analysis on
$N_t=6$ might underestimate the finite volume effects on finer lattices, since
due to lattice artefacts many of the staggered pion states are very heavy. Let
us also note here, that the volume independence in the transition region is an
unambiguous evidence for the crossover type of the transition.

In order to decrease lattice artefacts, we apply the tree-level improvement for
our thermodynamic observables (see Subsection \ref{methods_act}).  Figure
\ref{fig:rescale} illustrates at three temperature values ($T=132$, $167$ and
$223$ MeV) the effectiveness of this improvement procedure. We show both the
unimproved and the improved values of the trace anomaly for $N_t=6,8,10$ and
$12$ as a function of $1/N_t^2$. The lines are linear continuum extrapolations
using the three smallest lattice spacings.  As it can be seen in the continuum
limit both the unimproved and the improved observables converge to the same
value. The figure confirms the expectations, that lattice tree-level
improvement effectively reduces the cutoff effects. At all three temperatures
the unimproved observables have larger cutoff effects than the improved ones.
Actually, all the three values of $b_2(T)$, which indicate the remaining cutoff
effects after tree-level improvement (see Subsection \ref{methods_act}), differ
from zero with less than one standard deviation.

\subsection*{The $n_f=2+1$ flavor equation of state}
\begin{figure}[p]
\epsfig{file=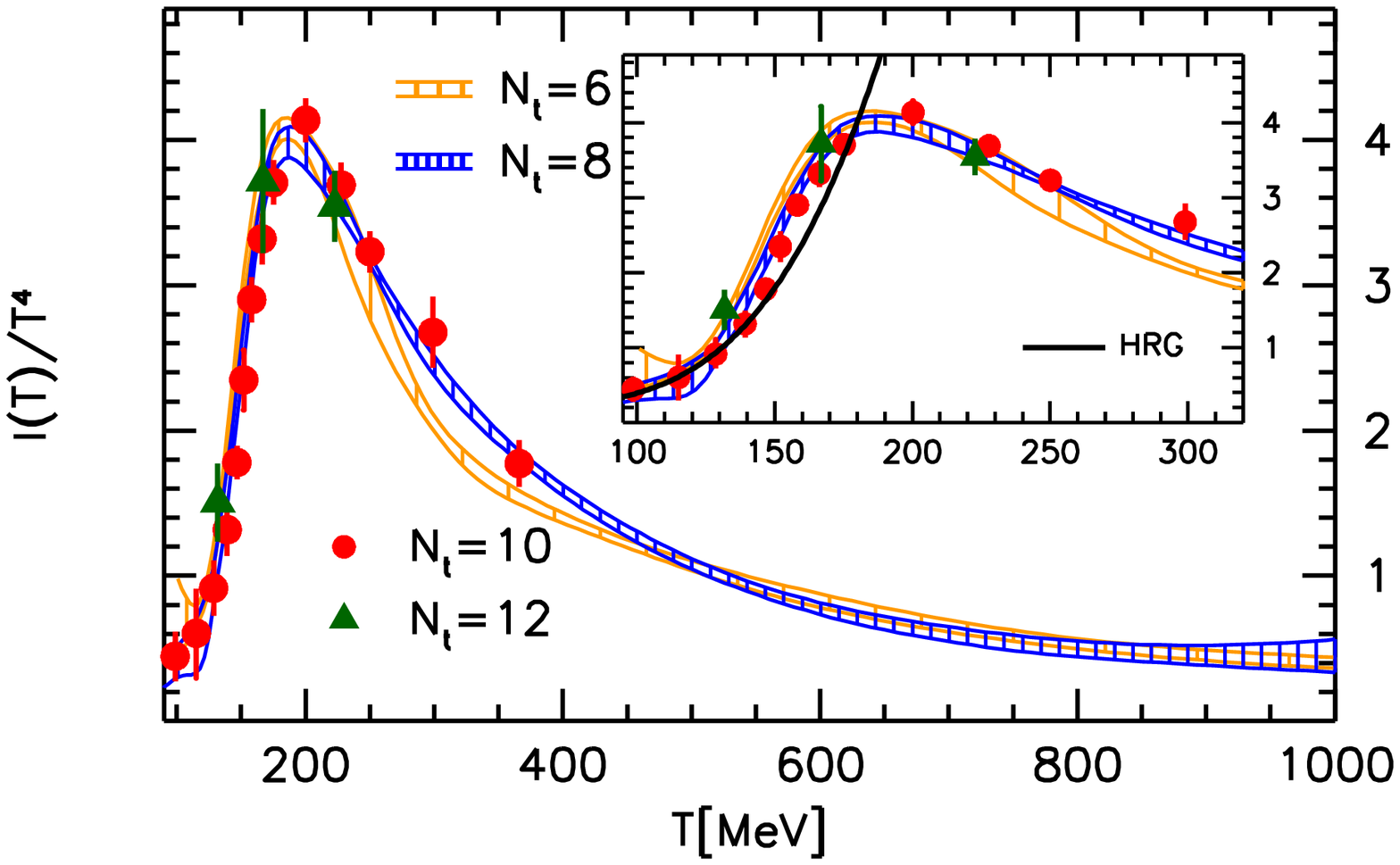,height=9cm,bb=18 360 592 718}
\caption{\label{fig:eos_I}
The trace anomaly $I=\epsilon-3p$ normalized by $T^4$ as a function of the temperature on $N_t=6,8,10$ and $12$ lattices.
}
\end{figure}
\begin{figure}[p]
\epsfig{file=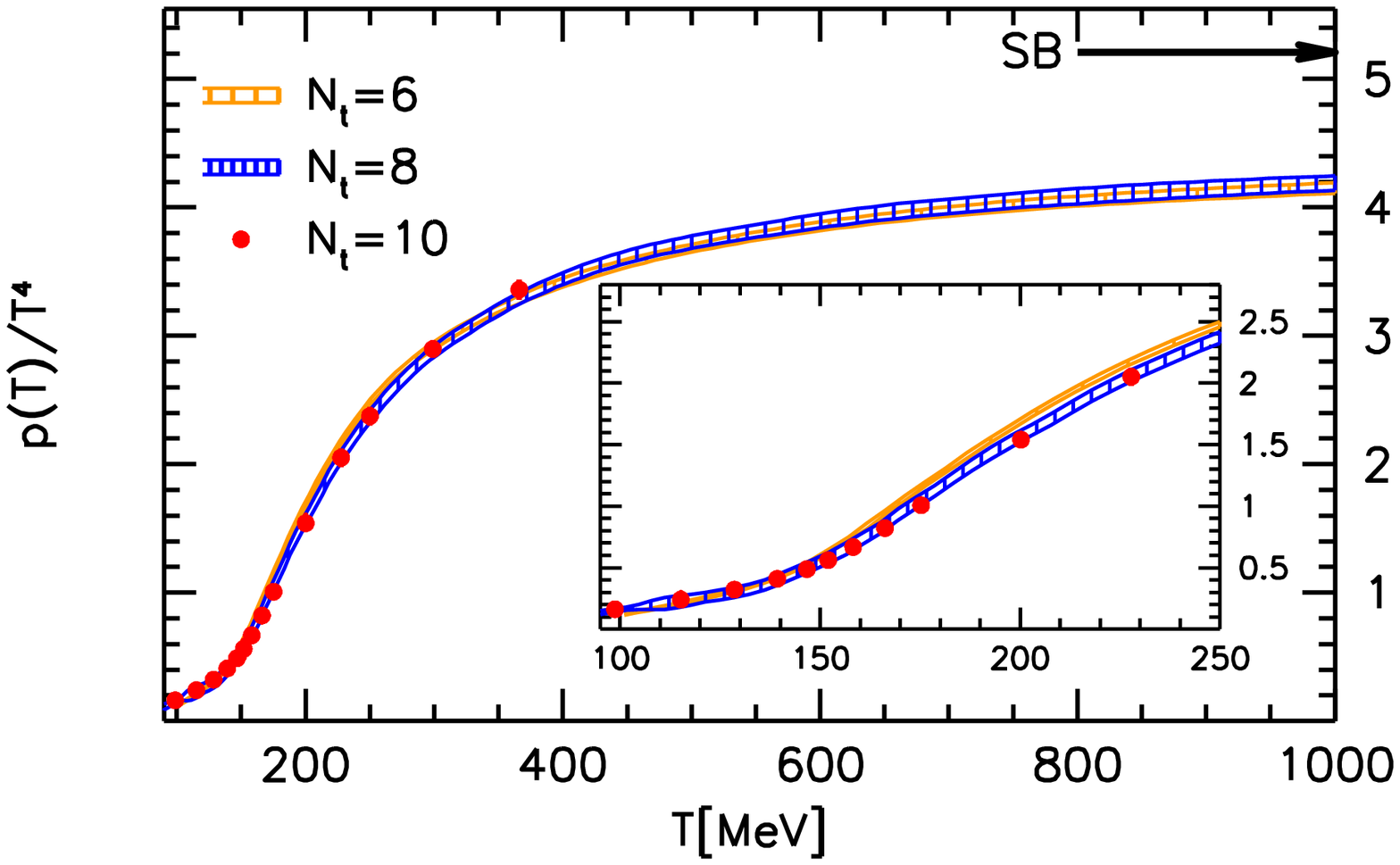,height=9cm,bb=18 360 592 718}
\caption{\label{fig:eos_p}
The pressure normalized by $T^4$
as a function of the temperature
on $N_t=6,8$ and $10$ lattices.
The Stefan-Boltzmann limit $p_{SB}(T) \approx 5.209 \cdot T^4$ is indicated by an arrow. For our highest temperature
$T=1000$ MeV the pressure is almost 20\% below this limit.
}
\end{figure}
On Figure \ref{fig:eos_I} we show the trace anomaly of QCD with $n_f=2+1$
flavor dynamical quarks as a function of the temperature. We have results at
four different lattice spacings. As it can be seen on the figure, our results
show essentially no dependence on the lattice spacing, they all lie on top of
each other.  Only the coarsest $N_t=6$ lattice shows some deviations
around $\sim 300$ MeV. On the same figure, we also provide a zoom of the
transition region. Here we also show the results from the HRG
model: a good agreement with the lattice results is found up to $T\sim 140$ MeV.
One characteristic temperature of the crossover transition
can be defined as the inflection point of the trace anomaly. This and other
characteristic features of the trace anomaly are the following:
\begin{center}
\begin{tabular}{|l|r|}
\hline
Inflection point of $I(T)/T^4$&152(4) MeV\\
Maximum value of $I(T)/T^4$&4.1(1)\\
$T$ at the maximum of $I(T)/T^4$&191(5) MeV\\
\hline
\end{tabular}
\end{center}

On Figure \ref{fig:eos_p} we show the main result of our paper: the pressure of
QCD with $n_f=2+1$ flavor dynamical quarks as a function of the temperature.
We have results at three different lattice spacings. The $N_t=6$ and $N_t=8$
are in the temperature range from 100 up to 1000 MeV. The results on $N_t=10$
are in the range from 100 up to 365 MeV. As we have already discussed in
Subsection \ref{methods_int}, the zero point of the pressure is set by Equation
(\ref{eq:nullpt}). From this condition we get a nonzero pressure even at our
smallest temperature $T=100$ MeV, when the physical values of the quark masses
are used. This value is approximately two third of the value suggested by the
HRG model. The origin of this difference cannot be clarified at the moment.
One expects that the lattice artefacts are considerably larger at low
temperatures, than what one estimates from the difference of $N_t=6,8$ and $10$
results. This is quite reasonable, since even our finest lattice at $T=100$ MeV
has $\sim 0.2$ fm lattice spacing, which is far from the regime, where lattice
results starts to scale. On the other hand, this discrepancy might point to the
failure of the HRG model. In order to be on the safe side we consider the size
of this unexplained difference as an estimate of our systematic uncertainty in
the low temperature regime.

\begin{figure}[p]
\epsfig{file=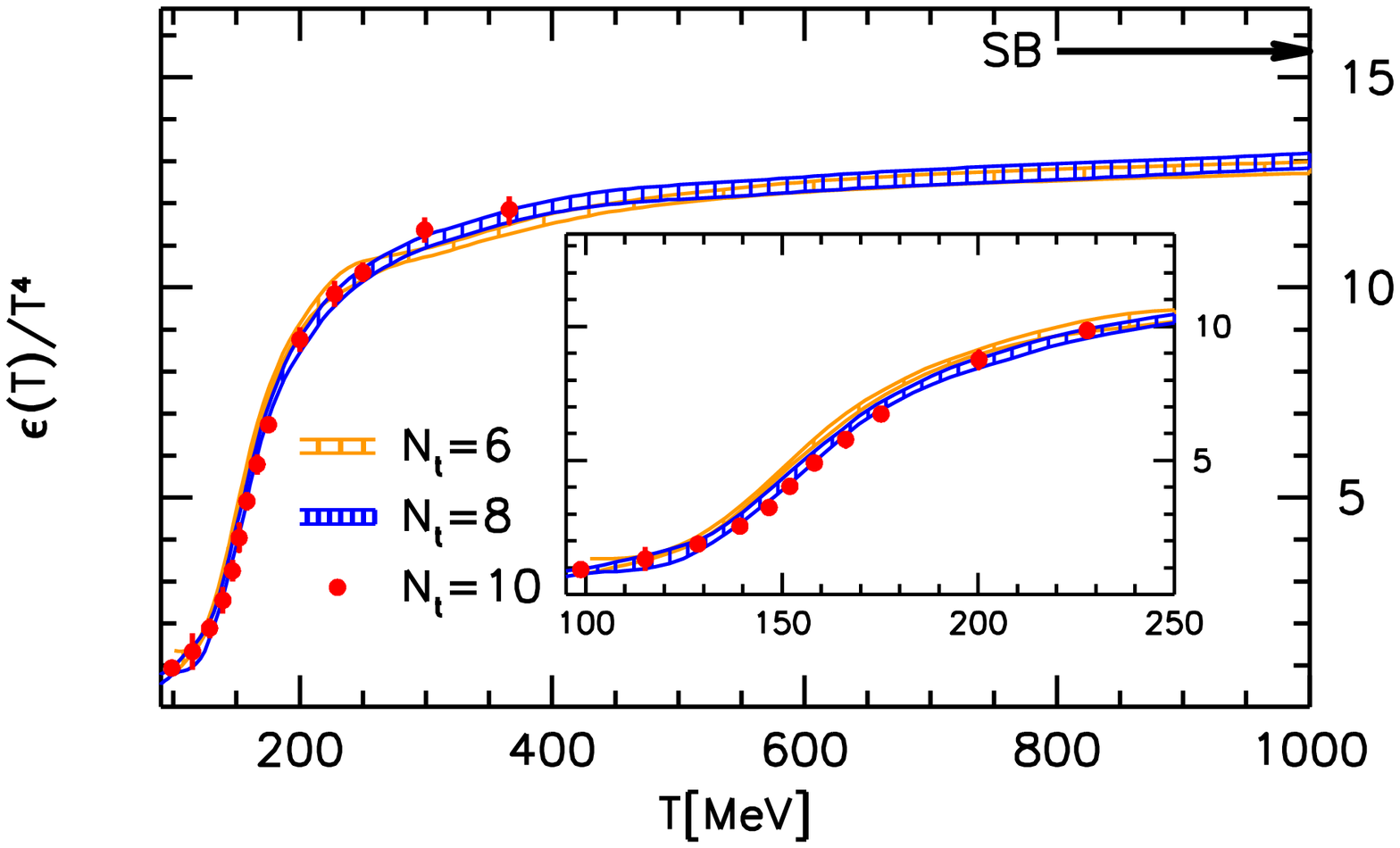,height=9cm,bb=18 360 592 718}
\caption{\label{fig:eos_e}
The energy density normalized by $T^4$
as a function of the temperature
on $N_t=6,8$ and $10$ lattices.
The Stefan-Boltzmann limit $\epsilon_{SB}= 3p_{SB}$ is indicated by an arrow.
}
\end{figure}
\begin{figure}[p]
\epsfig{file=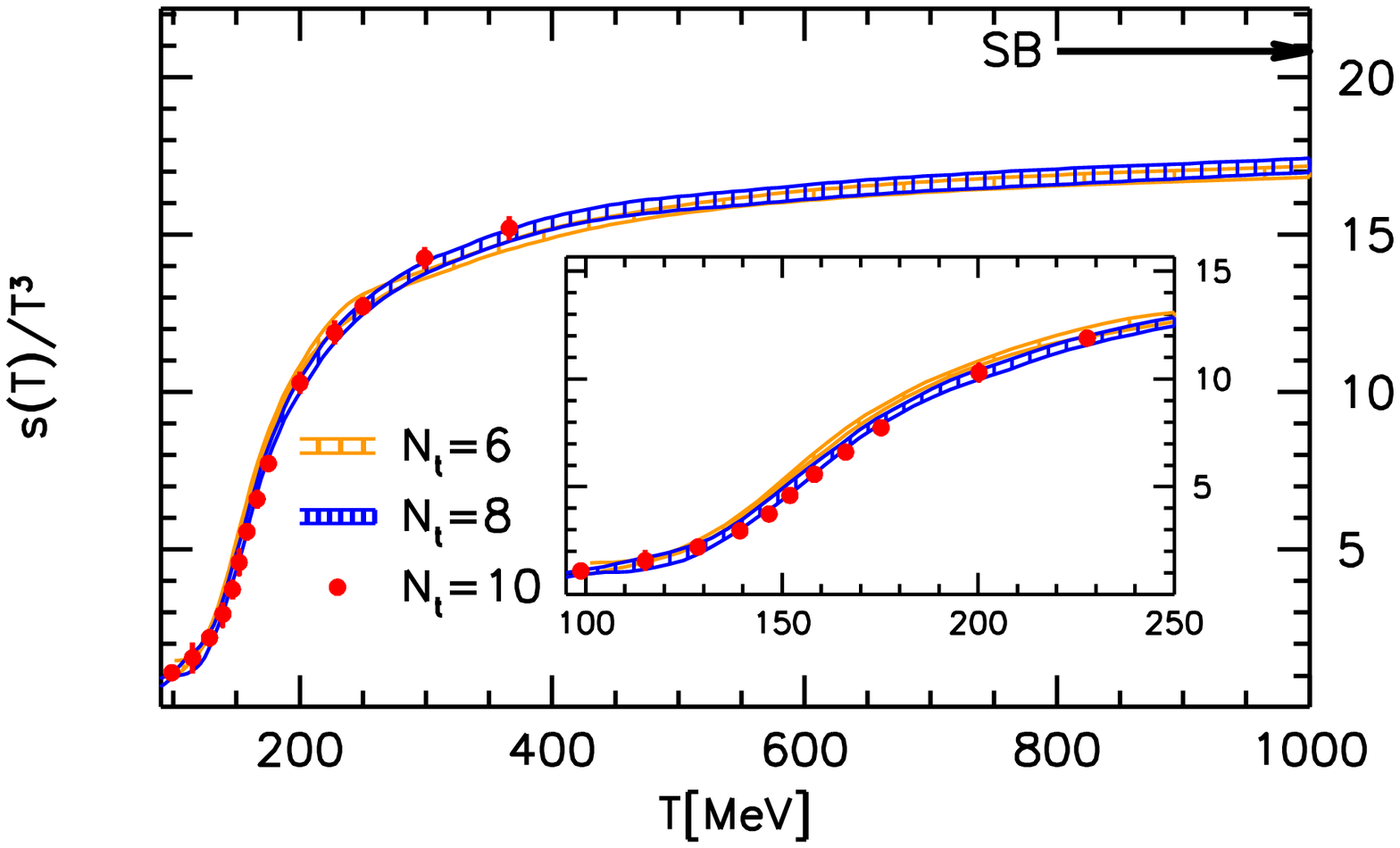,height=9cm,bb=18 360 592 718}
\caption{\label{fig:eos_s}
The entropy density normalized by $T^3$
as a function of the temperature
on $N_t=6,8$ and $10$ lattices.
The Stefan-Boltzmann limit $s_{SB}= 4p_{SB}/T$ is indicated by an arrow.
}
\end{figure}
\begin{figure}[p]
\epsfig{file=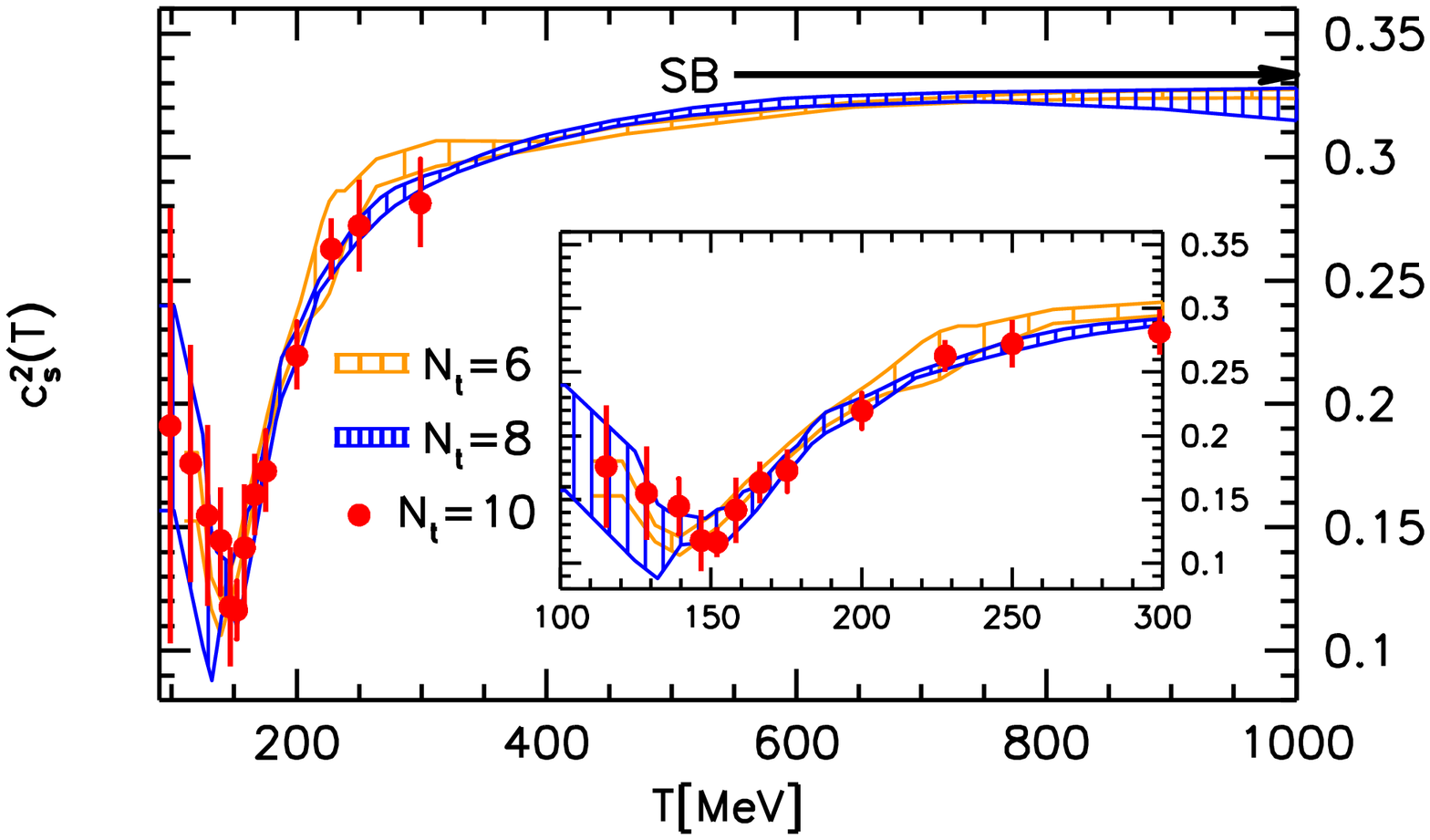,height=9cm,bb=18 360 592 718}
\caption{\label{fig:eos_cs}
The squared of the speed of sound 
as a function of the temperature on $N_t=6,8$ and $10$ lattices. The Stefan-Boltzmann limit is $c_{s,SB}^2=1/3$ 
indicated by an arrow.
}
\end{figure}
\begin{figure}[p]
\epsfig{file=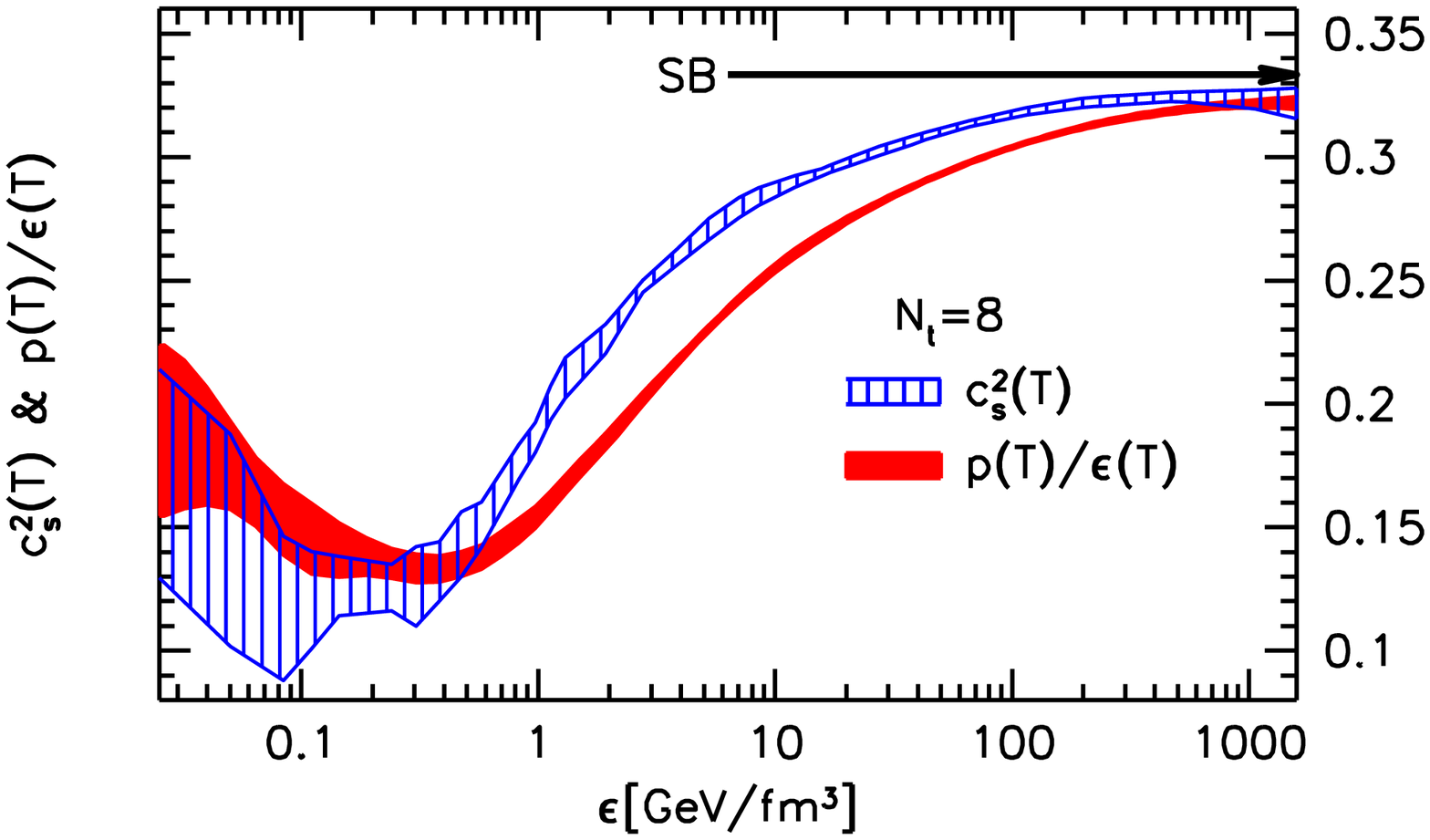,height=9cm,bb=18 360 592 718}
\caption{\label{fig:eos_pe}
The speed of sound and $p/\epsilon$ as a function of the energy density on $N_t=8$ lattices. 
The Stefan-Boltzmann limit is indicated by an arrow.
}
\end{figure}

On Figures \ref{fig:eos_e} and \ref{fig:eos_s} we show the energy density and
the entropy density, on Figures \ref{fig:eos_cs} and \ref{fig:eos_pe} the speed
of sound and $p/\epsilon$ are shown. On the latter we plot the quantities as
functions of the energy density. One can also read off the characteristic
points of these curves, for convenience we tabulate the results here:
\begin{center}
\begin{tabular}{|l|r|}
\hline
Minimum value of $c_s^2(T)$&0.133(5)\\
$T$ at the minimum of $c_s^2(T)$&145(5) MeV\\
${\epsilon}$ at the minimum of $c_s^2(T)$&0.20(4) GeV/fm$^3$\\
\hline
Minimum value of $p/\epsilon$&0.145(4)\\
$T$ at the minimum of $p/\epsilon$&159(5) MeV\\
${\epsilon}$ at the minimum of $p/\epsilon$&0.44(5) GeV/fm$^3$\\
\hline
\end{tabular}
\end{center}

In Appendix \ref{app_table} we tabulate the pressure, the trace anomaly and the
speed of sound for the $N_t=6,8$ and $10$ lattices. We also provide a continuum
estimate\footnote{For a rigorous continuum extrapolation one would need
$N_t=12$ for the entire temperature region.} for these quantities: we take the
average of the data at the smallest two lattice spacings and as an error we
assign either the half-difference of the two or the statistical error depending on
whichever is larger. As we have already explained, for low temperatures the
lattice result for the pressure is significantly smaller than the prediction of
the HRG model: at $T=100$ MeV the lattice result is $p(T)/T^4=0.16(4)$, whereas
the model prediction is $p(T)/T^4=0.27$.  Therefore in the continuum estimate
of the pressure we shift the central values of the lattice results up by the
half of this difference ($0.06$) and this shift is then considered as a
systematic error in the entire temperature range.

\begin{figure}[p]
\epsfig{file=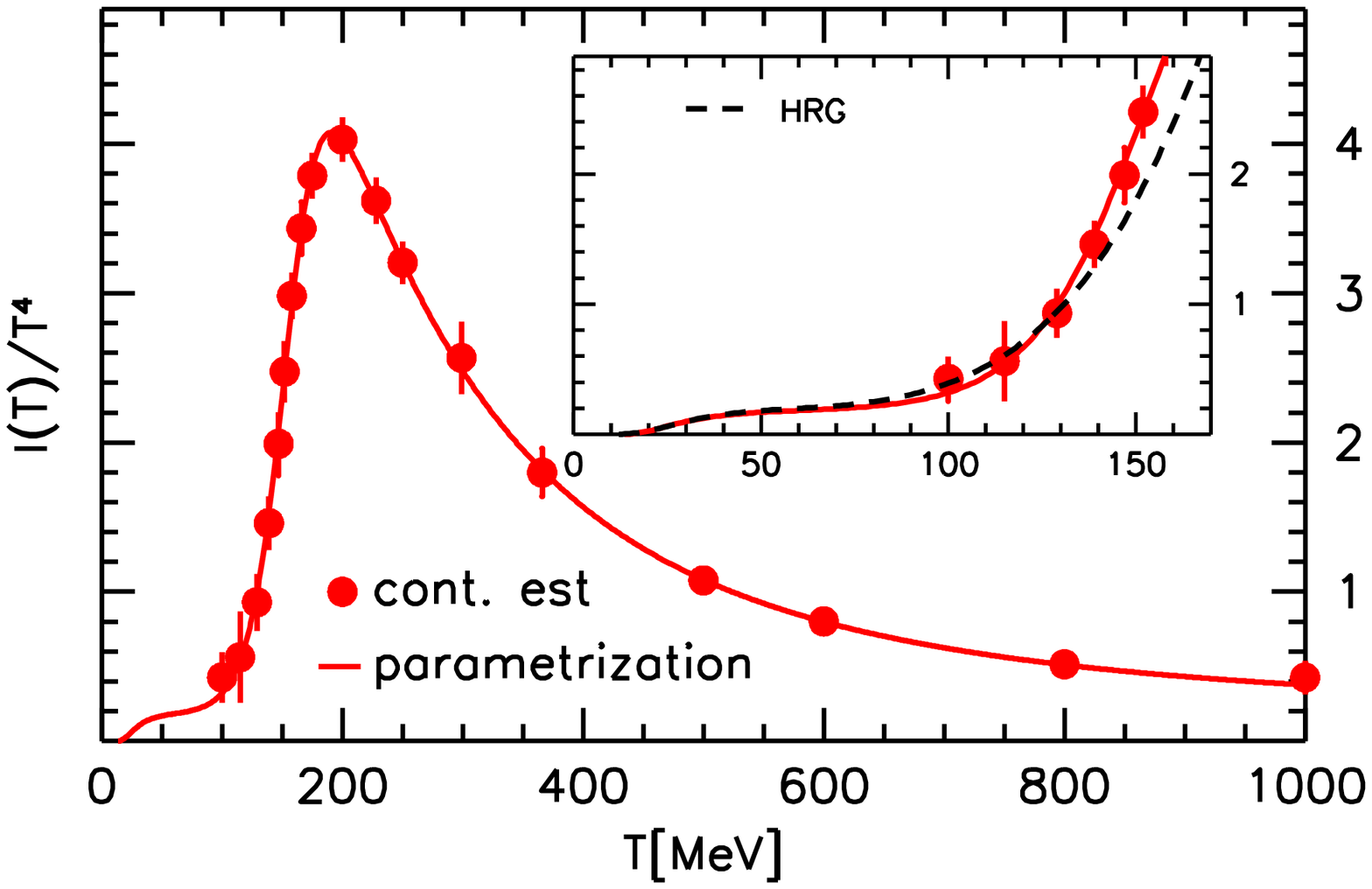,height=9cm,bb=18 360 592 718}
\caption{\label{fig:par}
Continuum estimate for the trace anomaly normalized by $T^4$ together with the parametrization
of Equation (\ref{eq:par}) using the $n_f=2+1$ parameters from Table \ref{tab:par}.
}
\end{figure}
For the reader's convenience we also give a global parametrization of the trace
anomaly as a function of the temperature. We took the following fit function:
\be
\label{eq:par}
\frac{I(T)}{T^4}= 
\exp(-h_1/t - h_2/t^2)\cdot \left( h_0 + \frac{f_0\cdot [\tanh(f_1\cdot t+f_2)+1]}{1+g_1\cdot t+g_2\cdot t^2} \right),
\ee
where the dimensionless $t$ variable is defined as $t=T/(200 {\rm MeV})$. 
The parameters can be found in Table \ref{tab:par}.
This function reproduces the continuum estimate for the normalized trace anomaly
in the entire temperature range $T=100\ldots 1000$ MeV.
The $\{f_0,f_1,f_2\}$ parameters describe the steep rise of the trace anomaly
in the transition region, whereas the $\{g_1,g_2\}$ parametrize the decrease
for higher temperatures. The parametrization also approximates the HRG model
prediction for $T<100$ MeV, this is described by the $\{h_0,h_1,h_2\}$
parameters. For these temperatures the difference in the trace anomaly between
the parametrization and the HRG model is less than $\Delta(I(T)/T^4) \le 0.07$.
From this parametrization the normalized pressure can be obtained by
the definite integral
\be
\frac{p(T)}{T^4}= \int_0^{T} \frac{dT}{T} \frac{I(T)}{T^4}.
\ee
The so obtained function
goes through the points of the continuum estimate of the pressure for
temperatures $T=100\ldots 1000$ MeV and for $T<100$ MeV the deviation from the
HRG prediction is less than $\Delta(p(T)/T^4) \le 0.02$. The parametrization together with our continuum estimate is 
shown on Figure \ref{fig:par}.
\TABLE[t]{
\begin{tabular}{|c||c|c|c|c|c|c|c|c|}
\hline
$n_f$ & $h_0$ & $h_1$ & $h_2$ & $f_0$ & $f_1$ & $f_2$ & $g_1$ & $g_2$ \\
\hline
$2+1$ & \multirow{2}{*}{0.1396} & \multirow{2}{*}{-0.1800} & \multirow{2}{*}{0.0350} & 2.76 & 6.79 & -5.29 & -0.47 & 1.04\\
\cline{1-1}\cline{5-9}
$2+1+1$ & & & & 5.59 & 7.34 & -5.60 & 1.42 & 0.50\\
\hline
\end{tabular}
\caption{\label{tab:par}
Parameters of the function in Equation (\ref{eq:par}) describing the trace anomaly in the $n_f=2+1$
and in the $n_f=2+1+1$ flavor cases.
}
}

\subsection*{Light quark mass dependence}
\begin{figure}[p]
\epsfig{file=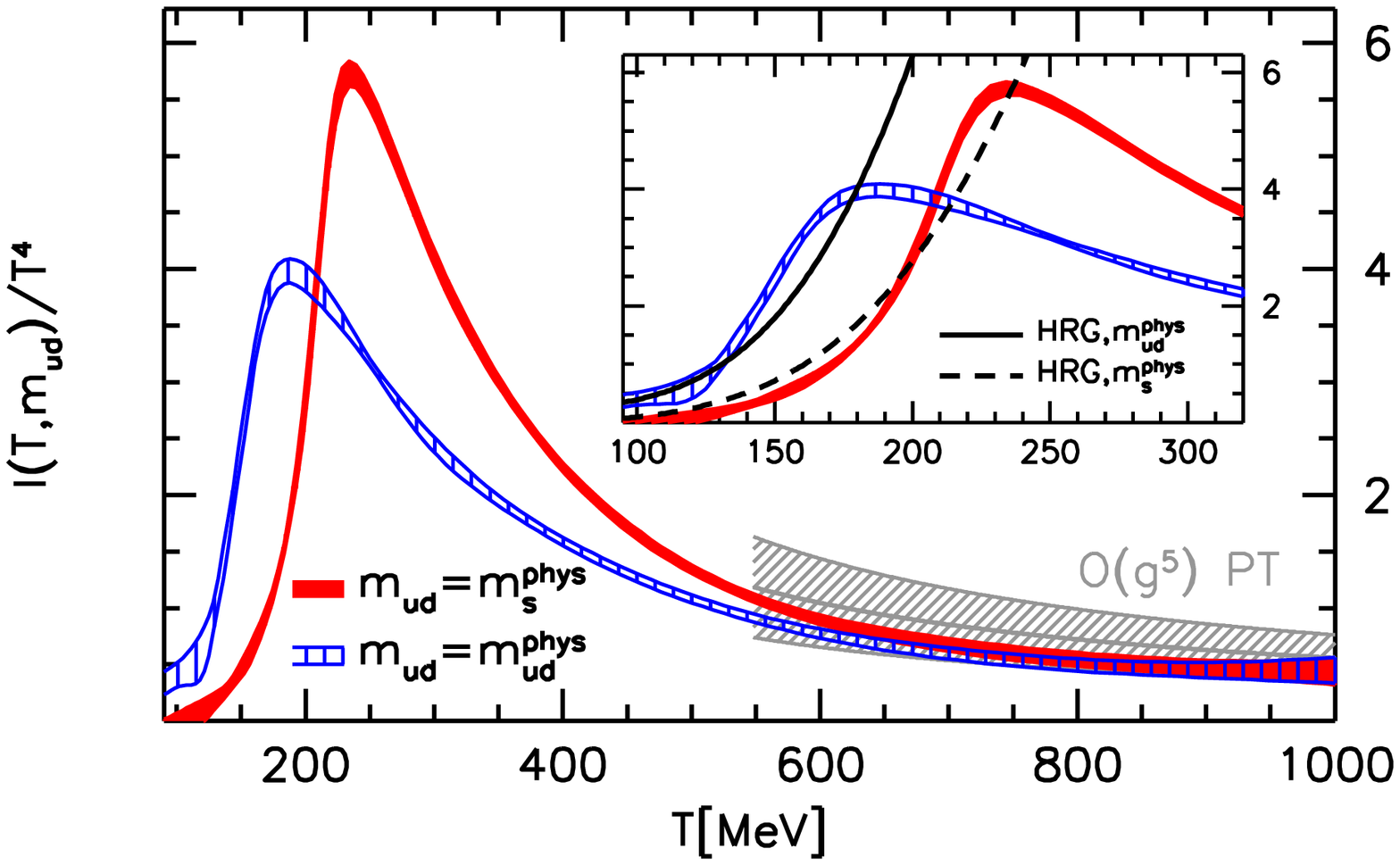,height=9cm,bb=18 360 592 718}
\caption{\label{fig:mdep}
The normalized trace anomaly for two different values of the light quark masses on $N_t=8$
lattices: the physical $m_{ud}= m_{ud}^{\rm phys}$ and the three degenerate
flavor $m_{ud}= m_s^{\rm phys}$ case.
}
\end{figure}

On Figure \ref{fig:mdep}, the trace anomaly for two different values of the
light quark masses is plotted: for the physical case, where $m_{ud}= m_{ud}^{\rm phys}$
and for the three degenerate flavor case, where $m_{ud}= m_s^{\rm phys}$. This latter case
corresponds to a pion mass of approximately $m_\pi \sim 720$ MeV.
The results are from
our $N_t=8$ lattices, this is the smallest lattice spacing, where we have the
complete mass dependence of the equation of state.  As it is expected, the peak
position of the trace anomaly is shifted towards higher temperature values for
larger quark masses. The position in the three degenerate flavor case is $\sim
25$\% larger than at the physical point.  The height also increases by about
$\sim 40$\%. When zooming into the transition region, we also show the
comparison with the HRG model. For low temperatures one finds a reasonable
agreement also in the heavy quark mass case. As it is expected, the dependence
on the quark masses vanishes as one goes to higher temperatures. Therefore it
sounds plausible to compare the result with that of the massless perturbation
theory. A good agreement can be observed with the highest order perturbative
calculation without nonperturbative input ($O(g^5)$, see Figure \ref{fig:mdep}).

In Appendix \ref{app_table} we tabulate the $N_t=8$ pressure and trace anomaly
for six different values of the quark mass ratio $R$.

\subsection*{Estimate for the $n_f=2+1+1$ flavor equation of state}
\begin{figure}[p]
\epsfig{file=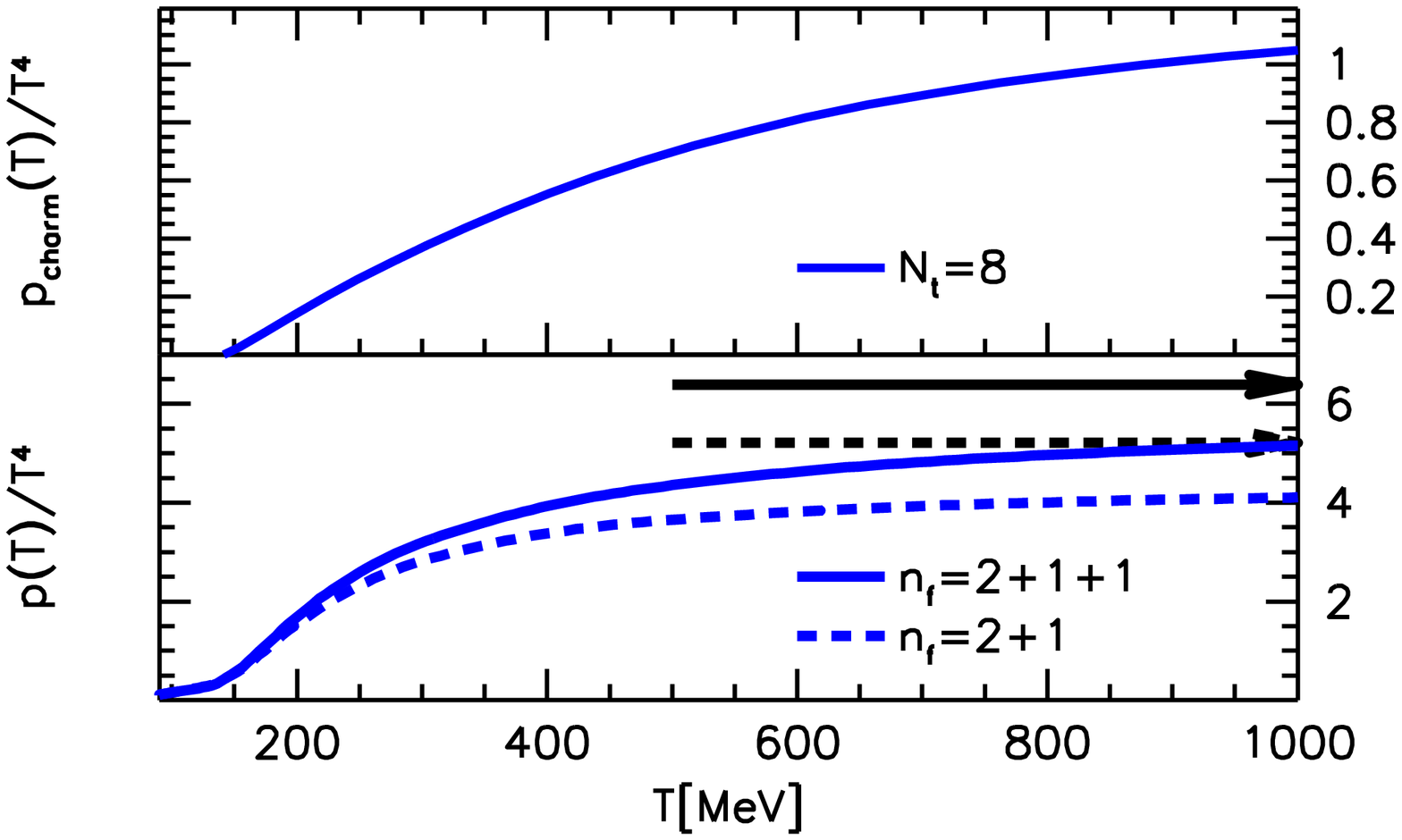,height=9cm,bb=18 360 592 718}
\caption{\label{fig:charm}
Upper figure: Contribution of the charm quark to the pressure on the $N_t=8$ lattices.
Lower figure: The pressure normalized by $T^4$ for
$n_f=2+1+1$ and $n_f=2+1$ flavors on $N_t=8$ lattices. The corresponding Stefan-Boltzmann limits are indicated
by arrows. The charm to strange quark mass ratio is $Q=11.85$ on this plot.
}
\end{figure}

While at low temperatures the equation of state only contains terms that
originate from the light quarks, in high energy processes charm quarks can also
be created from the vacuum, and they can also be present in the initial or
final states. One expects that the charm quark plays an important role already
above $T \gtrsim 2 \cdot T_c$ \cite{Laine:2006cp}. The inclusion of its
contribution is thus essential to determine the equation of state even in the
temperature range covered by our work. We estimated the contribution of the
charm quark on our $N_t=8$ lattices at several values of the charm to
strange quark mass ratio $Q$. According to a recent high-precision lattice
calculation \cite{Davies:2009ih} the physical value of $Q$ is $Q^{\rm
phys}=11.85(16)$. For this central value we show the contribution as a
function of temperature on the upper part of Figure \ref{fig:charm}. It is
non-zero already at temperatures $T\sim 200$ MeV.  The total $n_f=2+1+1$
pressure is compared to the already presented $n_f=2+1$ pressure on the lower
panel of Figure \ref{fig:charm}. Let us emphasize here again, that the estimate
for the charm contribution presented here suffers from two uncertainties: we
neglected the back-reaction of the charm quarks on the gauge field, moreover
due to the large mass of the charm large lattice artefacts are expected.

Using the parametrization in Equation (\ref{eq:par}) we fit the $n_f=2+1+1$ data
at the $N_t=8$ lattice spacing, the fit parameters can be found in Table
\ref{tab:par}. We tabulated the charm contribution together with results for
four other values of $Q$ in Appendix \ref{app_table}. From these, one also sees
that the uncertainty in the lattice determination of $Q^{\rm phys}$ has no
significant impact on the results.

\subsection*{Comparison with different fermion discretizations}
\begin{figure}
\begin{center}
\epsfig{file=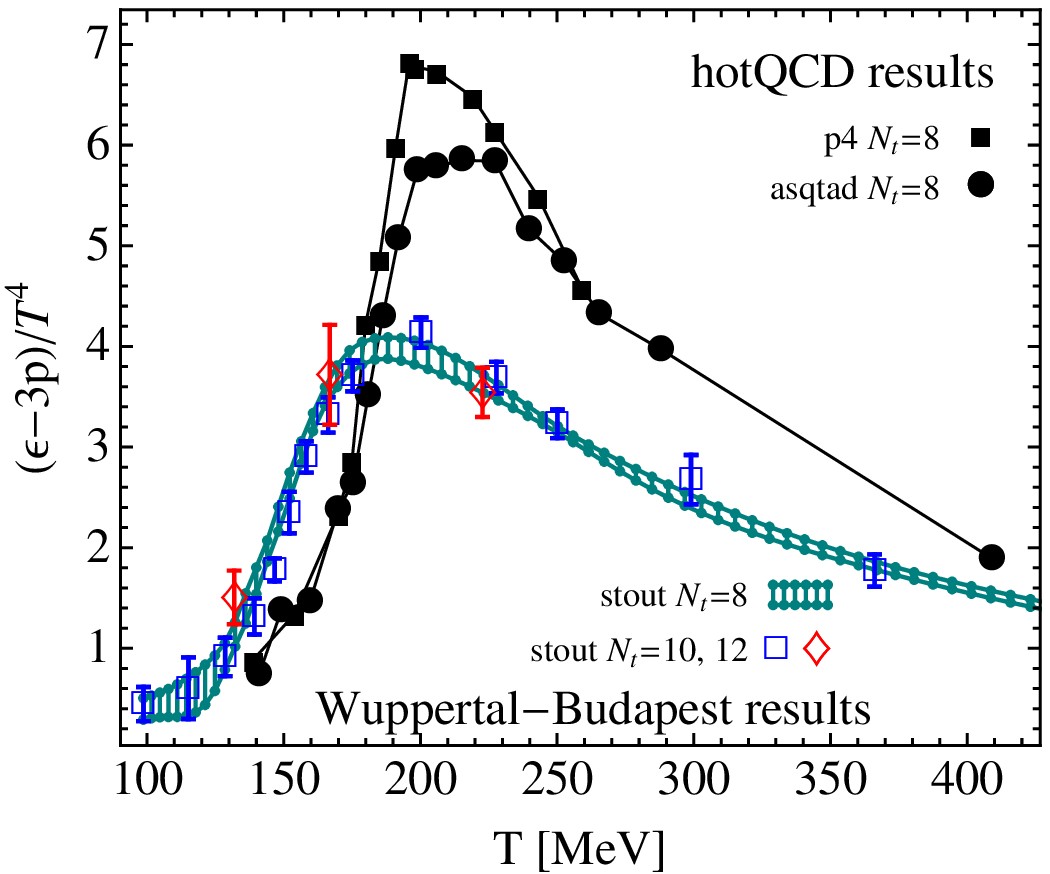,height=9cm}
\caption{\label{fig:cmp}
The normalized trace anomaly obtained in our study is compared to recent results from the ``hotQCD'' collaboration
\cite{Bazavov:2009zn,Cheng:2009zi}.
}
\end{center}
\end{figure}

As it was already discussed in Reference \cite{Aoki:2009sc}, there is a
disagreement between the results of current large scale thermodynamical
calculations. The main difference can be described by a $\sim$ 20-30 MeV shift
in the temperature.  This means, that the transition temperatures are
different: the temperature values that we obtain are smaller by this amount
than the values of the ``hotQCD'' collaboration. References
\cite{Huovinen:2009yb} and \cite{Borsanyi:2010bp} presented a possible
explanation for this problem: the more severe discretization artefacts of the
``asqtad'' and ``p4fat'' actions used by the ``hotQCD'' collaboration lead to larger
transition temperatures.

It is interesting to look for this discrepancy in the equation of state as
well. On Figure \ref{fig:cmp} we compare the trace anomaly obtained in this
study with the trace anomaly of the ``hotQCD'' collaboration.  We plot the
$N_t=8$ data using the ``p4fat'' and ``asqtad'' actions, which we took from
References \cite{Bazavov:2009zn} and \cite{Cheng:2009zi}.  As it can be clearly
seen, the upward going branch and the peak position are located at $\sim$ 20 MeV
higher temperatures in the simulations of the ``hotQCD'' group.  This is the same
phenomenon as the one, which was already reported for many other quantities in
Reference \cite{Aoki:2009sc}. We also see, that the peak height is about 50\%
larger in the ``hotQCD'' case.

%% file: summary.tex
In this paper we determined the equation of state of QCD by means of lattice
simulations. Results for the $n_f=2+1$ flavor pressure, trace anomaly, energy
and entropy densities and for the speed of sound were presented on figures and
in tables. We gave a simple parametrization for the trace anomaly. Moreover, the
light quark mass dependence of the equation of state was investigated
quantitatively. We also addressed the $n_f=2+1+1$ flavor equation of state: an
estimate of the charm quark contribution to the thermodynamic potential at
the partially quenched level was given and a non-negligible effect down to
temperatures of $T\sim 200$ MeV was found.

The results were obtained by carrying out lattice simulations at four different
lattice spacings $N_t=6,8,10$ and $12$ in the temperature range
$T=100\dots1000$ MeV. We developed and tested a new method to obtain the Lines
of Constant Physics for these high temperature values. We also proposed a new
method to get the pressure from quantities measured on the lattice: instead of
the standard numerical integration procedure we fit a many parameter function
to the lattice data. In order to reduce the lattice artefacts we applied
tree-level improvement for all of the thermodynamical observables. We found
that there is no difference in the results at the three finest lattice
spacings. This shows that the lattice discretization errors are not significant
and the continuum limit can be reliably taken. For low temperatures however,
the apparent absence of the discretization artefacts might be misleading.
Using the Hadron Resonance Gas model we estimated the discretization error of
the lattice results for low temperatures. We ruled out the existence of
significant finite size effects by comparing our results to a data set with
double linear box size. 

A comparison with the results of the ``hotQCD'' collaboration was made. The
already reported problem, ie. the ``hotQCD'' results tend to give $20-30$ MeV
larger temperature values, is also present in the equation of state.
Additionally we found that the peak of the trace anomaly is about 50\%
larger in the ``hotQCD'' calculations.

In order to provide a more precise calculation of the $n_f=2+1+1$ flavor
equation of state, there is still room for improvements. First, one has to
relax the uncontrolled partial quenched approximation and introduce the charm
quark dynamically. Then, the lattice spacing has to be further reduced to
ensure a good resolution of the charmed excitations. Finally, improved fermion
actions are increasingly important when studying heavy quark degrees of
freedom. A further obvious direction of improvement of the present work is to
abandon the uncontrolled ``rooting'' issue of the staggered discretization
and solve the same problem with e.g. Wilson-Clover fermions.

%% file: app_spline.tex
In this Appendix we describe the method, which was used to obtain the pressure
as a function of the parameters $\beta$ and $R$ using its measured derivatives
(Equations (\ref{eq:dbeta}) and (\ref{eq:dR})). Further details can be found in
Reference \cite{Gergospline}.

Let us first construct a two dimensional bicubic spline function $P(\beta,R)$.
We start from a grid of node points $\{\beta_k,R_l\}$ with $0\le k<K$ and $0\le
l<L$. Upon this grid $P(\beta,R)$ is unambiguously determined by the values
$p_{kl}$ that it takes at these node points. The spline coefficients can be
written compactly as $C_{ab,kl}$ with $0\le a<4$ and $0\le b <4$ being the
indices for the appropriate powers of $\beta$ and $R$, and $k$ and $l$ indicate
the corresponding grid square $[\beta_k,\beta_{k+1}]\times[R_l,R_{l+1}]$. There
are $16\cdot(K-1)\cdot(L-1)$ coefficients altogether. The spline function $P$
and its first and second derivatives are continuous along the grid lines
$\{\beta_k,\cdot\}$ and $\{\cdot,R_l\}$, and the second derivatives go to zero
at the two ends. These constraints and the condition 
\be
P(\beta_k,R_l) = p_{kl}
\ee
constitute a linear system of equations for $C_{ab,kl}$, that can be solved to give
\be
C_{ab,kl} = \sum\limits_{k'=0}^{K-1}\sum\limits_{l'=0}^{L-1} X_{ab,kl;k'l'}\; p_{k'l'}
\label{eq:c}
\ee
with $X$ being a matrix of size $16\cdot(K-1)\cdot(L-1)\times K \cdot L$. 

Next we determine the spline coefficients by fitting the derivatives of the
$P(\beta,R)$ function to the measured derivatives $D_\beta$ and $D_R$ in
Equations (\ref{eq:dbeta}) and (\ref{eq:dR}). Since the two derivatives are
determined using the same configurations at a given value of the bare
parameters, their correlation has to be taken into account in the $\chi^2$
fitting. The $\chi^2$ function will be a quadratic function of the values
$p_{kl}$ and therefore it can be minimized by solving a $K\cdot L$ dimensional
system of linear equations:
\be
\sum_{k'=0}^{K-1}\sum_{l'=0}^{L-1}M_{kl,k'l'} p_{k'l'} = V_{kl}.
\label{eq:sys}
\ee
Here the matrix $M$ and the vector $V$ can be calculated using the matrix $X$
and the values of the measured derivatives $D_\beta$ and $D_R$. The system of linear
equations in Equation (\ref{eq:sys}) can be solved for $p_{kl}$ and with it the
spline coefficients are also determined through Equation (\ref{eq:c}). These
unambiguously determine the whole pressure surface $P(\beta, R)$. The solution
is of course indeterminate up to an overall constant, which does not influence
the gradient of the surface. 

The coordinates of the node points in the $R$-direction were set to be halfway
between the measurement points ranging from $R_l=1.0 \ldots 28.15$ in order to
ensure the stability of the fit. The coordinates in the $\beta$-direction on
the other hand were varied randomly around a ``stable'' fit to estimate the
systematic error of this fitting procedure. The number of node points were also
varied. This analysis shows that for the pressure the systematic error related
to our integration procedure is around or less than the statistical one. For
the trace anomaly, this error is larger for some temperature values. We added
the statistical and systematic errors in quadrature.

%% file: app_table.tex
\TABLE{
\begin{tabular}{|c|c|c|c|c|c|c|}
\hline
\multicolumn{7}{|c|}{$p(T)/T^4$}\\
\hline
\multirow{2}{*}{T[MeV]}&\multicolumn{6}{|c|}{$m_\pi$[MeV]}\\
\cline{2-7}
 &$716$&$414$&$271$&$191$&$160$&$135$\\
\hline
100 &{\bf 0}&0.05(1)&0.10(1)&0.13(1)&0.14(1)&0.16(1)\\
115 &-0.02(3)&0.03(6)&0.13(4)&0.14(6)&0.23(6)&0.23(5)\\
129 &0.02(4)&0.11(4)&0.19(4)&0.25(4)&0.28(4)&0.30(4)\\
134 &0.03(4)&0.12(4)&0.22(4)&0.29(4)&0.32(4)&0.34(4)\\
139 &0.04(4)&0.15(4)&0.25(4)&0.33(4)&0.37(4)&0.40(4)\\
143 &0.04(4)&0.17(4)&0.30(4)&0.38(4)&0.42(4)&0.45(4)\\
147 &0.05(4)&0.20(4)&0.34(4)&0.43(4)&0.47(4)&0.51(4)\\
152 &0.07(4)&0.24(4)&0.39(4)&0.50(4)&0.54(4)&0.58(4)\\
158 &0.09(4)&0.30(4)&0.48(5)&0.59(5)&0.64(5)&0.68(5)\\
162 &0.11(4)&0.34(4)&0.53(5)&0.67(5)&0.72(5)&0.76(5)\\
166 &0.13(4)&0.39(5)&0.60(4)&0.75(4)&0.80(4)&0.85(4)\\
170 &0.15(5)&0.43(5)&0.68(5)&0.84(5)&0.90(5)&0.94(5)\\
175 &0.18(5)&0.51(5)&0.80(5)&0.95(4)&1.01(5)&1.05(5)\\
185 &0.25(5)&0.72(5)&1.03(4)&1.19(4)&1.24(4)&1.27(4)\\
200 &0.43(5)&1.06(5)&1.38(5)&1.51(5)&1.54(5)&1.57(5)\\
215 &0.68(5)&1.43(5)&1.70(4)&1.80(5)&1.83(5)&1.85(5)\\
228 &0.98(4)&1.72(5)&1.95(5)&2.03(5)&2.05(5)&2.06(5)\\
250 &1.51(5)&2.13(5)&2.29(5)&2.35(5)&2.36(5)&2.37(5)\\
275 &2.00(4)&2.49(5)&2.61(5)&2.64(5)&2.65(5)&2.66(5)\\
299 &2.39(5)&2.75(5)&2.84(5)&2.87(5)&2.87(5)&2.87(5)\\
330 &2.74(4)&3.00(4)&3.06(5)&3.08(5)&3.08(5)&3.08(5)\\
366 &3.06(5)&3.24(5)&3.28(5)&3.29(5)&3.29(5)&3.29(5)\\
400 &3.28(5)&3.42(5)&3.44(5)&3.44(5)&3.45(5)&3.45(5)\\
450 &3.52(5)&3.60(5)&3.61(5)&3.61(5)&3.62(5)&3.62(5)\\
500 &3.67(5)&3.73(5)&3.73(5)&3.73(5)&3.73(5)&3.73(5)\\
600 &3.87(5)&3.89(5)&3.89(5)&3.88(6)&3.90(5)&3.90(5)\\
800 &4.08(6)&4.09(6)&4.09(6)&4.09(6)&4.09(6)&4.09(6)\\
1000 &4.19(6)&4.19(6)&4.19(6)&4.19(6)&4.19(6)&4.19(6)\\
\hline
\end{tabular}
\caption{\label{tab:mdep_p}
$N_t=8$ results for the normalized pressure $p(T)/T^4$ at different pion masses
described by the quark mass ratio $R=m_s^{\rm phys}/m_{ud}$. The pion
masses correspond to $R=1, 3, 7, 14, 20$ and $28.15$.  The zero point of the
pressure was set at $R=1$, $T=100$ MeV.
}
}

\newpage

\TABLE{
\begin{tabular}{|c|c|c|c|c|c|c|}
\hline
\multicolumn{7}{|c|}{$I(T)/T^4$}\\
\hline
\multirow{2}{*}{T[MeV]}&\multicolumn{6}{|c|}{$m_\pi$[MeV]}\\
\cline{2-7}
 &$716$&$414$&$271$&$191$&$160$&$135$\\
\hline
100 &-0.04(11)&0.26(09)&0.41(07)&0.35(09)&0.37(10)&0.41(11)\\
115 &0.05(11)&0.41(10)&0.63(08)&0.38(16)&0.47(18)&0.52(19)\\
129 &0.17(07)&0.57(08)&0.88(09)&0.80(11)&0.91(12)&0.95(13)\\
134 &0.22(06)&0.64(07)&0.98(09)&1.11(11)&1.22(11)&1.25(13)\\
139 &0.28(05)&0.74(06)&1.13(09)&1.39(13)&1.54(13)&1.60(14)\\
143 &0.34(05)&0.84(06)&1.29(09)&1.60(11)&1.80(11)&1.89(11)\\
147 &0.40(06)&1.00(07)&1.50(09)&1.86(12)&2.08(12)&2.20(12)\\
152 &0.49(08)&1.26(08)&1.85(09)&2.28(12)&2.49(13)&2.60(13)\\
158 &0.62(08)&1.66(09)&2.35(09)&2.87(10)&2.99(10)&3.06(10)\\
162 &0.72(08)&1.97(09)&2.69(09)&3.23(09)&3.29(09)&3.33(09)\\
166 &0.83(07)&2.30(09)&3.04(10)&3.55(09)&3.55(11)&3.55(09)\\
170 &0.96(06)&2.65(09)&3.37(10)&3.80(10)&3.75(12)&3.72(11)\\
175 &1.14(05)&3.10(08)&3.75(10)&4.01(11)&3.92(13)&3.87(11)\\
185 &1.62(05)&3.97(08)&4.32(09)&4.18(10)&4.06(12)&3.98(10)\\
200 &2.83(14)&4.90(13)&4.64(13)&4.12(12)&4.00(13)&3.92(12)\\
215 &4.65(13)&5.06(12)&4.33(11)&3.93(10)&3.81(12)&3.75(11)\\
228 &5.63(13)&4.76(10)&3.96(09)&3.71(08)&3.60(08)&3.55(08)\\
250 &5.51(09)&4.08(06)&3.52(06)&3.31(03)&3.22(03)&3.18(03)\\
275 &4.82(07)&3.44(06)&3.03(05)&2.87(05)&2.80(05)&2.78(05)\\
299 &4.13(07)&2.96(07)&2.62(06)&2.51(06)&2.47(06)&2.46(07)\\
330 &3.40(06)&2.49(06)&2.22(06)&2.15(05)&2.14(06)&2.13(06)\\
366 &2.73(04)&2.07(04)&1.88(04)&1.82(04)&1.83(04)&1.82(04)\\
400 &2.25(04)&1.75(03)&1.63(03)&1.58(04)&1.60(04)&1.59(04)\\
450 &1.72(04)&1.40(04)&1.33(04)&1.30(04)&1.31(04)&1.31(04)\\
500 &1.35(04)&1.14(04)&1.10(04)&1.09(04)&1.08(04)&1.08(04)\\
600 &0.90(04)&0.80(04)&0.80(04)&0.80(05)&0.76(04)&0.77(04)\\
800 &0.55(06)&0.51(05)&0.51(06)&0.52(06)&0.49(05)&0.50(05)\\
1000 &0.43(11)&0.44(10)&0.43(11)&0.40(13)&0.48(12)&0.45(11)\\
\hline
\end{tabular}
\caption{\label{tab:mdep_I}
$N_t=8$ results for the normalized trace anomaly $I(T)/T^4$ at different pion
masses described by the quark mass ratio $R=m_s^{\rm phys}/m_{ud}$. The
pion masses correspond to $R=1, 3, 7, 14, 20$ and $28.15$.
}
}

\newpage

\begin{landscape}
\TABLE{
\begin{tabular}{|c||c|c|c||c|c|c||c|c|c||c|c|c||}
\hline
\multirow{2}{*}{T[MeV]} & \multicolumn{3}{|c||}{$N_t=6$}& \multicolumn{3}{c||}{$N_t=8$}& \multicolumn{3}{c||}{$N_t=10$} & \multicolumn{3}{c||}{cont. estimate} \\
\cline{2-13}
& $p/T^4$ & $I/T^4$ & $c_s^2$& $p/T^4$ & $I/T^4$ & $c_s^2$& $p/T^4$ & $I/T^4$ & $c_s^2$ & $p/T^4$ & $I/T^4$ & $c_s^2$\\
\hline
100 &0.12(0)&0.71(29)&0.19(9)&0.16(1)&0.41(11)&0.20(2)&0.16(4)&0.45(17)&0.19(9)&0.22(4)&0.43(17)&0.19(9)\\
115 &0.21(0)&0.72(7)&0.19(2)&0.23(5)&0.52(19)&0.19(3)&0.24(6)&0.60(31)&0.18(5)&0.29(6)&0.56(31)&0.18(5)\\
129 &0.32(1)&1.16(8)&0.13(0)&0.30(4)&0.95(13)&0.12(3)&0.32(6)&0.91(19)&0.15(4)&0.37(6)&0.93(19)&0.14(4)\\
139 &0.42(1)&1.89(9)&0.11(0)&0.40(4)&1.60(14)&0.12(1)&0.41(6)&1.32(18)&0.14(2)&0.46(6)&1.46(18)&0.13(2)\\
147 &0.54(1)&2.58(8)&0.12(0)&0.51(4)&2.20(12)&0.13(1)&0.49(7)&1.78(11)&0.12(2)&0.55(7)&1.99(21)&0.12(2)\\
152 &0.63(2)&2.97(9)&0.13(0)&0.58(4)&2.60(13)&0.13(2)&0.56(7)&2.35(21)&0.12(1)&0.63(7)&2.47(21)&0.12(2)\\
158 &0.75(2)&3.37(10)&0.15(1)&0.68(5)&3.06(10)&0.14(1)&0.67(7)&2.90(16)&0.14(3)&0.73(7)&2.98(16)&0.14(3)\\
166 &0.94(2)&3.76(11)&0.17(0)&0.85(4)&3.55(9)&0.15(0)&0.82(7)&3.32(18)&0.16(2)&0.89(7)&3.43(18)&0.16(2)\\
175 &1.15(2)&4.00(9)&0.19(0)&1.05(5)&3.87(11)&0.18(0)&1.01(6)&3.71(15)&0.17(2)&1.08(6)&3.79(15)&0.18(2)\\
200 &1.69(2)&3.98(8)&0.23(0)&1.57(5)&3.92(12)&0.22(0)&1.54(6)&4.14(15)&0.22(1)&1.61(6)&4.03(15)&0.22(1)\\
228 &2.18(2)&3.47(19)&0.27(2)&2.06(5)&3.55(8)&0.25(0)&2.05(7)&3.69(16)&0.26(1)&2.11(7)&3.62(16)&0.26(1)\\
250 &2.48(2)&2.97(20)&0.29(0)&2.37(5)&3.18(3)&0.27(0)&2.38(7)&3.23(14)&0.27(2)&2.43(7)&3.20(14)&0.27(2)\\
299 &2.92(3)&2.06(5)&0.30(0)&2.87(5)&2.46(7)&0.29(0)&2.90(7)&2.68(24)&0.28(2)&2.94(7)&2.57(24)&0.29(2)\\
366 &3.28(3)&1.54(5)&0.30(0)&3.29(5)&1.82(4)&0.30(0)&3.36(8)&1.77(16)&0.33(3)&3.38(8)&1.80(16)&0.32(3)\\
500 &3.68(3)&1.07(4)&0.31(0)&3.73(5)&1.08(4)&0.32(0)&-&-&-&3.76(5)&1.08(4)&0.32(0)\\
600 &3.85(3)&0.83(5)&0.32(0)&3.90(5)&0.77(4)&0.32(0)&-&-&-&3.93(5)&0.80(5)&0.32(0)\\
800 &4.05(4)&0.53(3)&0.32(0)&4.09(6)&0.50(5)&0.32(0)&-&-&-&4.12(6)&0.51(5)&0.32(0)\\
1000 &4.15(4)&0.40(4)&0.33(0)&4.19(6)&0.45(11)&0.32(0)&-&-&-&4.23(6)&0.43(11)&0.32(0)\\
\hline
\end{tabular}
\caption{\label{tab:cont}
Lattice data for the pressure, trace anomaly and the speed of sound as
functions of the temperature. We also give a continuum estimate based on the
mean of the finest two discretizations at a given temperature. In the continuum
estimate of $p/T^4$ there is an additional systematic uncertainty of $0.06$, which 
is not included in the table.
}
}
\end{landscape}

\newpage

\TABLE{
\begin{tabular}{|c|c|c|c|c|c|}
\hline
\multicolumn{6}{|c|}{$p_{\rm charm}(T)/T^4$}\\
\hline
\multirow{2}{*}{$T$[MeV]} & \multicolumn{5}{|c|}{$Q$} \\
\cline{2-6}
 & 10.75 & 11.85 & 12.5 & 16.0 & 20.0 \\
\hline
147 &0.01(0)&0.01(0)&0.01(0)&0.01(0)&0.01(0)\\
152 &0.02(0)&0.02(0)&0.02(0)&0.02(0)&0.01(0)\\
158 &0.04(0)&0.04(0)&0.04(0)&0.03(0)&0.02(0)\\
162 &0.05(0)&0.05(0)&0.05(0)&0.04(0)&0.03(0)\\
166 &0.06(0)&0.06(0)&0.06(0)&0.04(0)&0.03(0)\\
170 &0.07(0)&0.07(0)&0.07(0)&0.05(0)&0.04(0)\\
175 &0.09(0)&0.08(0)&0.08(0)&0.06(0)&0.05(0)\\
185 &0.12(0)&0.10(0)&0.10(0)&0.08(0)&0.06(0)\\
200 &0.16(0)&0.14(0)&0.14(0)&0.11(0)&0.08(0)\\
215 &0.20(0)&0.18(0)&0.18(0)&0.14(0)&0.10(0)\\
228 &0.23(0)&0.21(0)&0.20(0)&0.16(0)&0.12(0)\\
250 &0.28(0)&0.26(0)&0.25(1)&0.21(0)&0.15(0)\\
275 &0.34(0)&0.32(1)&0.31(1)&0.25(0)&0.19(1)\\
299 &0.39(0)&0.37(1)&0.35(1)&0.29(1)&0.22(1)\\
330 &0.46(0)&0.43(1)&0.41(1)&0.34(1)&0.26(1)\\
366 &0.53(1)&0.49(1)&0.48(1)&0.40(1)&0.31(2)\\
400 &0.58(1)&0.55(1)&0.53(1)&0.45(1)&0.35(2)\\
450 &0.66(1)&0.63(1)&0.61(1)&0.52(1)&0.41(2)\\
500 &0.73(1)&0.70(1)&0.68(1)&0.58(1)&0.46(2)\\
600 &0.84(1)&0.81(1)&0.79(1)&0.69(1)&0.56(1)\\
800 &0.98(1)&0.96(1)&0.95(1)&0.86(0)&0.73(1)\\
1000 &1.06(1)&1.05(1)&1.04(1)&0.97(0)&0.86(1)\\
\hline
\end{tabular} 
\caption{\label{tab:charm}
$N_t=8$ results for the charm contribution to the pressure $p_{\rm charm}(T)/T^4$ at different values
of the quark mass ratio $Q=m_c/m_s^{\rm phys}$. The value $Q=11.85$ corresponds
to a recent charm mass determination \cite{Davies:2009ih}. Note, that there are additional systematic
uncertainties related to the partial quenching approximation.
}
}